%% file: manuscript.tex
\documentclass[a4paper,10pt]{article} %

\input{packages}

\input{defs}
\begin{document}
\title{\textsc{Electro-mechanical wrinkling of soft dielectric films bonded to hyperelastic substrates}}

\author{\textsc{Bin Wu}$^1$ $\,\,\cdot$
    \textsc{Linghao Kong}$^1$ $\,\,\cdot$
    \textsc{Weiqiu Chen}$^{1,\,2,\,3,\,4}$ \\
    \textsc{Davide Riccobelli}$^5$ $\,\,\cdot$
    \textsc{Michel Destrade}$^2$\\
    \normalsize $^1$Key Laboratory of Soft Machines and Smart Devices of Zhejiang Province,\\
    \normalsize Department of Engineering Mechanics, and Soft Matter Research Center,\\
    \normalsize Zhejiang University, Hangzhou 310027, P.R. China.\\
    \normalsize $^2$School of Mathematical and Statistical Sciences, \\
    \normalsize University of Galway, University Road, Galway, Ireland.\\
    \normalsize $^3$Center for Soft Machines and Smart Devices, Huanjiang Laboratory, Zhuji 311816, P.R. China.\\
    \normalsize $^4$Faculty of Mechanical Engineering and Mechanics, Ningbo University, Ningbo 315211, P.R. China.\\
    \normalsize $^5$Mathematics Area, mathLab, SISSA – International School for Advanced Studies,\\
    \normalsize Via Bonomea 265, Trieste 34136, TS, Italy.}
\date{\today}

\maketitle

\begin{abstract}
    Active control of wrinkling in soft film-substrate composites using electric fields is a critical challenge in tunable material systems.
    Here, we investigate the electro-mechanical instability of a soft dielectric film bonded to a hyperelastic substrate, revealing the fundamental mechanisms that enable on-demand surface patterning. For the linearized stability analysis, we use the Stroh formalism and the surface impedance method to obtain exact and sixth-order approximate bifurcation equations that signal the onset of wrinkles.
    We derive the explicit bifurcation equations giving the critical stretch and critical voltage for wrinkling, as well as the corresponding critical wavenumber.
    We look at scenarios where the voltage is kept constant and the stretch changes, and vice versa.
    We provide the thresholds of the shear modulus ratio $r_{\rm c}^0$ or pre-stretch $\lambda_{\rm c}^0$ below which the film-substrate system wrinkles mechanically, prior to the application of a voltage.
    These predictions offer theoretical guidance for practical structural design, as the shear modulus ratio $r$ and/or the pre-stretch $\lambda$ can be chosen to be slightly greater than $r_{\rm c}^0$ and/or $\lambda_{\rm c}^0$, so that the film-substrate system wrinkles with a small applied voltage.
    Finally, we simulate the full nonlinear behavior using the Finite Element method (\texttt{FEniCS}) to validate our formulas and conduct a post-buckling analysis. This work advances the fundamental understanding of electro-mechanical wrinkling instabilities in soft material systems. By enabling active control of surface morphologies via applied electric fields, our findings open new avenues for adaptive technologies in soft robotics, flexible electronics, smart surfaces, and bioinspired systems.
\end{abstract}

\section{Introduction}

Surface wrinkling of soft materials and biological tissues is ubiquitous in nature and engineering \citep{li2012mechanics,tan2020bioinspired}, typically {\color{black}occurring} when a soft substrate coated with a stiffer film is loaded mechanically beyond a critical threshold \citep{liu2024surface}.
In biology, countless wrinkling morphologies appear, such as the folds of brain matter \citep{griffiths2009atlas,fernandez2016cerebral,balbi2020mechanics,riccobelli2020surface}, the track of oesophageal mucosa \citep{li2011surface}, and the wrinkles of skin \citep{autumn2002evidence,zhao2020multi}.
In engineering, the wrinkling of film-substrate systems can be harnessed to design specific patterns and alter optical properties \citep{li2017harnessing}, probe the surface characteristics of materials \citep{stafford2004buckling}, reduce effective surface tension \citep{lee2021dependence}, design novel nonlithographic phase masks \citep{zhao2020path},  {\color{black}and} help design novel flexible sensors \citep{wang2022flexible,lee2022surface,Yin_2024ultra-soft}, etc.
Therefore, exploring the buckling and post-buckling regimes of film-substrate systems helps us understand and further control multiple pattern formations. However, the purely mechanical actuation of wrinkling/creasing in film-substrate systems does not allow for efficient active control of such surface patterns \citep{Psarra_2017Two}.

The emergence of soft smart materials provides a great opportunity for applied research {\color{black}on} film-substrate systems. Specifically, soft dielectric elastomers (DEs), which deform significantly under an external electric field \citep{pelrine1998electrostriction}, offer the advantages of extensive actuation strains \citep{Li_2013Giant}, fast response \citep{chen2019controlled}, and low elastic modulus \citep{Shian_2015Dielectric}, {\color{black}enabling their use} in artificial muscles \citep{Brochu2009Advances}, electrical energy storage devices \citep{li2018high}, sensors \citep{pelrine1998electrostriction,lee2022surface,Yin_2024ultra-soft}, grippers \citep{Shian_2015Dielectric}, and soft robots \citep{Liang_2020Comparative,Guo_2021Review}.

The coupling of Maxwell's equations of electricity with those of continuum mechanics complicates the study of film-substrate instabilities.
For pure elastic film-substrate systems, early studies concentrated on analyzing linearized stability in the neighborhood of large contractions \citep{Shield_1994Buckling, Ogden_1996The,cai1999imperfection}, and were followed by advanced explorations {\color{black}of} wrinkling, post-{\color{black}buckling}, semi-{\color{black}analytic methods,} and finite element simulations \citep{Cai_2000Exact, Cao_2011wrinkles, cao2012Wrinkling,Hutchinson_2013the, Fu2015Buckling, Cheewaruangroj_2019Pattern,Alawiye_2019Revisiting, Alawiye_2020Revisiting, liu2024surface}.
Regarding pure DE systems, many works {\color{black}have focused on} instability \citep{Bertoldi_2011Instabilities, Fu_2018Localized, su2018wrinkles, Su_2019Finite, Su_2023Tunable, Su_2024Electromechanical, zhu2024voltage, Si_2025Voltage}.

\begin{figure}[t!]
    \centering
    \setlength{\abovecaptionskip}{5pt}
    \includegraphics[width=0.9\textwidth]{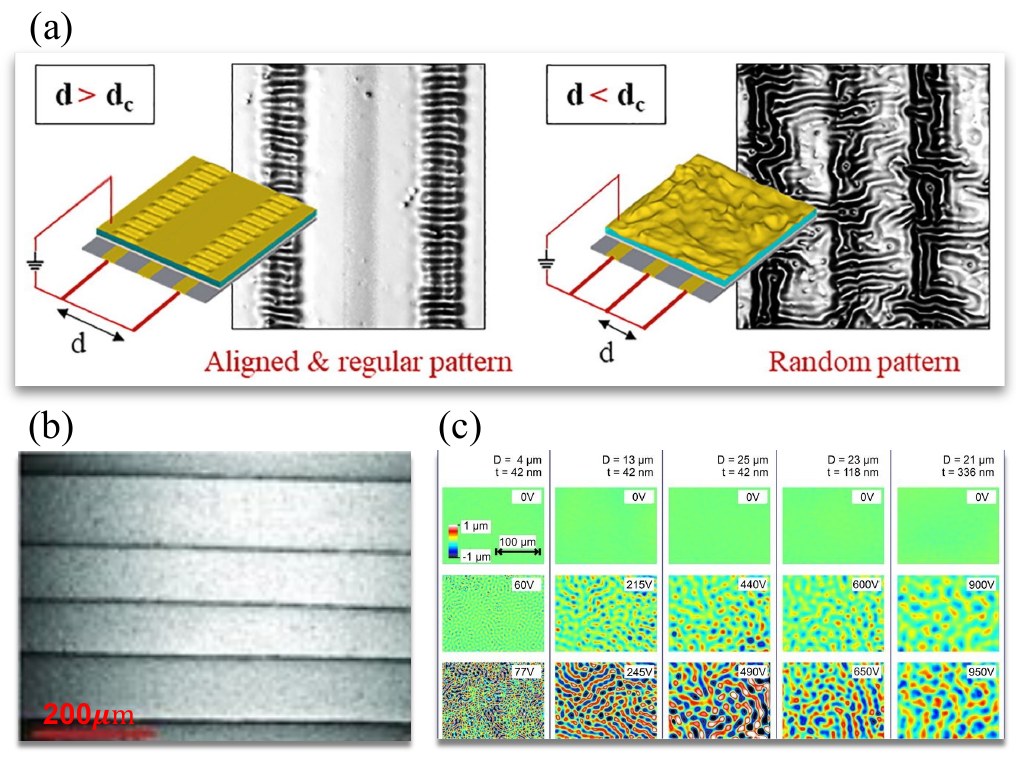}
    \caption{Examples of electrically induced surface instabilities in soft polymer films. These patterns are formed not on a soft substrate as studied in this work, but on a rigid one, which limits the system's deformability {\color{black}but still} illustrates key principles of electro-responsive wrinkling.
        (a) Wrinkle patterns can be switched from regular to random by tuning the spacing $d$ between {\color{black}the} underlying electrodes \citep{lin2020electro};
        (b) Highly aligned parallel lines can be formed by applying a voltage to a uniaxially pre-stretched film \citep{wang2012dynamic};
        (c) On films without pre-stretch, increasing {\color{black}the} voltage causes a flat surface to buckle into random wrinkles and craters above a critical threshold \citep{van_den_Ende_2013}.}
    \label{Fig-intro}
\end{figure}

\cite{Kofod_2003Actuation,Wang_2011Electro, wang2011creasing} and \cite{wang2013creasing} carried out a series of experiments on pre-stretched elastic dielectrics bonded to \emph{rigid} substrates and subjected to {\color{black}high voltages}, leading to localized creasing-like instabilities (see Fig.~\ref{Fig-intro}), which were later studied theoretically and numerically by \cite{Hutchinson_2021Surface} and \cite{Landis_2022Formation}.
In those cases, however, the rigidity of the substrates imposes significant limitations on applications.
Systems comprising a dielectric film bonded to a soft substrate have also been studied: for example, \cite{Su_2020Pattern} investigated the bending deformation of a dielectric-elastic bilayer in response to a voltage; \cite{Almamo2024Axisymmetric} studied the axisymmetric vibrations of a dielectric-elastic tubular bilayer system{\color{black};} and \cite{Sriram_2024Data} used a data-driven approach to model the onset of wrinkling in composite {\color{black}DE} bilayer structures subjected to combined electro-mechanical loading conditions.

The conclusion of this survey is that the potential instabilities and pattern formation of {\color{black}DE films} bonded to a soft hyperelastic substrate (Fig.~\ref{Fig1}) are yet to be analyzed theoretically and numerically.
This work combines the advantages of DEs (a type of smart material) and film-substrate systems to investigate the stability of a soft dielectric film bonded to a hyperelastic substrate under a plane-strain mechanical load and a uniform transverse electric field (or voltage) (Fig.~\ref{Fig1}(b)).
We work within the framework of nonlinear electro-elasticity theory and the associated linearized incremental theory developed by \cite{VERMA1966small} and \cite{dorfmann2014nonlinear}.
To overcome the complexity of conventional displacement-based methods, we use the Stroh formulation and the surface impedance method \citep{su2018wrinkles} to derive exact solutions and approximate explicit expressions {\color{black}for} the bifurcation equations.
In addition, we use the finite element method based on \texttt{FEniCS} to conduct {\color{black}a} wrinkling analysis of the DE film-substrate system, and the results are compared with the theoretical solutions. Finally, {\color{black}a} post-buckling analysis of the DE film-substrate system is also conducted.

Our results show that the onset of wrinkling in a soft dielectric film-substrate system can be actively tuned by {\color{black}electro-mechanical} loading, provided the material parameters are chosen appropriately. In particular, we find explicit formulas for the critical stretch and {\color{black}critical} voltage at which wrinkles emerge, along with the corresponding wrinkle wavelength.

We look at two loading path scenarios: first, the {\color{black}applied} voltage is fixed at the equilibrium value {\color{black}at which} there is no applied traction, and a mechanical load is applied; second, the stretch is fixed at a given contractile or extensile level, and the voltage is increased.
The analytical bifurcation results reveal a threshold stiffness ratio and pre-stretch (denoted $r_{\rm c}^0$ and $\lambda_{\rm c}^0$) below which the film-substrate system wrinkles under purely mechanical compression, even with no voltage applied. Unsurprisingly, this {\color{black}corresponds to} the bifurcation criterion for the compression of non-coupled, hyperelastic systems. Above these thresholds, however, the film remains flat until a sufficient voltage triggers the instability, in the contractile as well as the extensile regimes.
This behavior provides a practical design guideline: by selecting the substrate-to-film stiffness ratio and pre-stretch just above $r_{\rm c}^0$ and $\lambda_{\rm c}^0$, one can ensure the {\color{black}system} stays smooth under mechanical load and then wrinkles \emph{on demand} with a small applied voltage (note that if the system is pre-stretched in extension, a potentially {\color{black}high} voltage is required for wrinkling).
We verify these predictions through finite element simulations, which not only confirm the accuracy of the critical stretch and voltage estimates, but also capture the post-buckling evolution of the wrinkle patterns, including the potential development of period-doubling and period-tripling patterns.
Importantly, our stability analysis indicates that the wrinkle formation is a supercritical bifurcation {\color{black}in most cases}, meaning the pattern amplitude grows gradually from zero at the critical point (rather than jumping suddenly). This benign, progressive onset is favorable for applications because it ensures smooth and reliable actuation of the wrinkle pattern as conditions change.

Active control of surface instabilities via electric fields is a promising strategy in soft materials research.
This approach aligns with major efforts in morphing soft robotic components, flexible and stretchable electronics, smart surface engineering, and bioinspired interfaces, where on-demand reconfigurability is a must.
Here, we combine theoretical modeling and finite element simulations to elucidate the electro-mechanical instability of a soft dielectric film bonded to a hyperelastic substrate, addressing this timely challenge from {\color{black}both} fundamental and computational perspectives.
Our findings not only shed light on the mechanics of electrically induced wrinkling but also demonstrate how electric fields can be {\color{black}exploited} to actively tune surface patterns, {\color{black}thereby} broadening the design space for functional soft materials.

\section{Materials and methods}
\label{section2}

\subsection{Setup}

\begin{figure}[t!]
    \centering
    \setlength{\abovecaptionskip}{5pt}
    \includegraphics[width=1.0\textwidth]{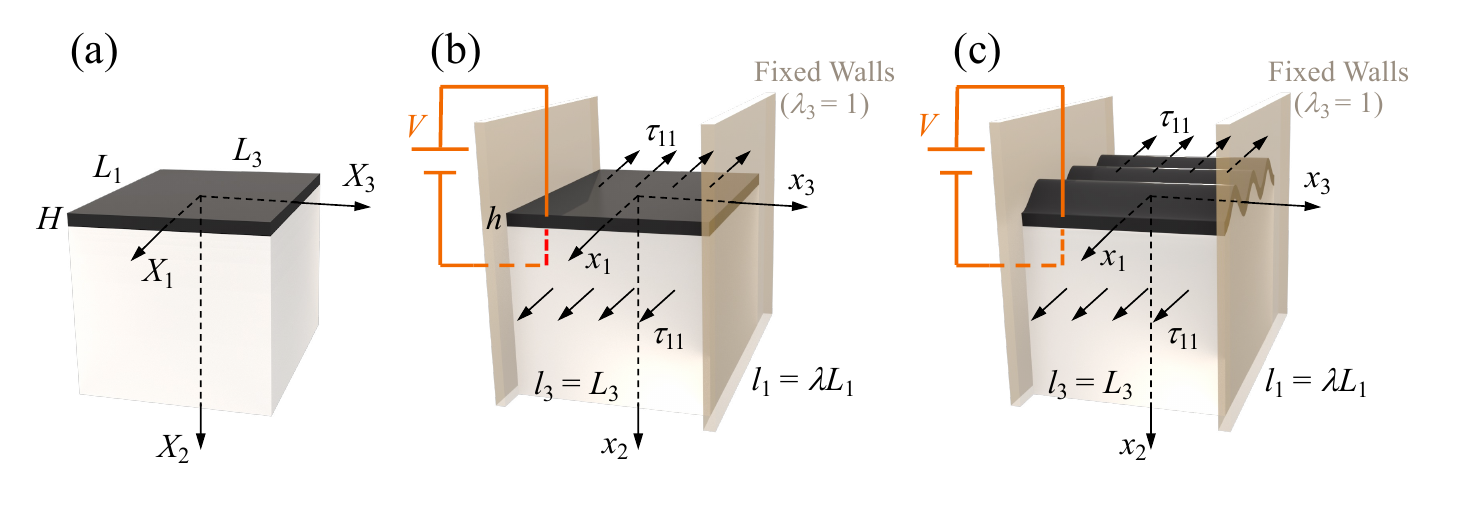}
    \caption{Schematic diagram of a soft dielectric film bonded to a hyperelastic substrate, confined between two lubricated rigid walls. (a) Initial {\color{black}undeformed} configuration; (b) Current {\color{black}deformed} configuration, prior to (c) the onset of {\color{black}wrinkles}.}
    \label{Fig1}
\end{figure}

The system consists of a semi-infinite elastic substrate and an elastic dielectric film glued on its surface.
In the initial {\color{black}undeformed} configuration, the substrate and the film occupy the $X_2>0$ and $-H<X_2<0$ regions, respectively, where $(X_1, X_2, X_3)$ are the {\color{black}Cartesian} coordinates and $H$ is the film thickness.
In the current {\color{black}deformed} configuration, the {\color{black}film-substrate} system is deformed homogeneously along the principal directions of stretch $x_i$ (parallel to the $X_i$), with corresponding stretch ratios $\lambda_i$.
Hence, the current film thickness is $h=\lambda_2 H$.
The film and substrate are perfectly bonded and both {\color{black}are} incompressible so that they undergo the same deformation and $\lambda_1\lambda_2\lambda_3=1$.

The deformation is due to the application of an {\color{black}electric} field inside the film, generated by the potential difference $V$ between two flexible electrodes coated on its top and bottom faces.
In general, a {\color{black}soft} dielectric film expands in its plane under a voltage.
Here, for simplicity and to make connections with known results, we focus on the {\color{black}plane-strain deformation} $\lambda_1=\lambda$, $\lambda_2=\lambda^{-1}$, $\lambda_3=1$, which is achieved by confining the system {\color{black}between} two fixed, lubricated, rigid plates, normal to the {\color{black}$x_3$}-direction, and applying {\color{black}external} forces in the {\color{black}$x_1$}-direction {\color{black}(see Fig.~\ref{Fig1}b;} note that in~\ref{AppeA}, we treat the general triaxial case).

To {\color{black}characterize} the materials, we choose the neo-Hookean model for the {\color{black}hyperelastic} substrate and the {\color{black}neo-Hookean} ideal dielectric model for the film, so that their \emph{total} free {\color{black}energy density functions} take the form
\begin{equation}
    \label{eq:energies}
    W_s = \tfrac{1}{2}\mu_s (\lambda^2+\lambda^{-2}-2),
    \qquad
    W_f = \tfrac{1}{2}\mu_f (\lambda^2+\lambda^{-2}-2) - \tfrac{1}{2}\varepsilon \lambda^2 \frac{V^2}{H^2},
\end{equation}
respectively, where the $\mu_i$ ($i=s,f$) are the initial shear moduli in the undeformed configuration, and $\varepsilon$ is the film's electric permittivity, which remains unaffected by the deformation.
The subscripts $s$ and $f$ refer to the physical quantities of the substrate and film, respectively.

When the film is under voltage $V$, the whole {\color{black}film-substrate} system deforms homogeneously.
Because the top surface at $x_2=-h$ is free of electro-elastic traction, the normal component of the total stress vanishes there. By continuity, it also vanishes at the perfectly bonded interface.
Then, the following \emph{total} Cauchy stress component along the $x_1$-direction is required to keep the plane-strain deformation in the film under voltage $V$ and stretch $\lambda$ \citep{su2018wrinkles}: $\sigma_f = \lambda W_f'(\lambda) = \mu_f(\lambda^2 - \lambda^{-2}) - \varepsilon \lambda^2(V/H)^2$.
It follows that at equilibrium, when no lateral traction is applied along $x_1$ (i.e., $\sigma_f =0$), the corresponding stretch $\lambda_0$ and voltage $V_0$ are linked as \citep{su2019tuning}
\begin{equation}
    \lambda_0 = (1-\bar E_0^2)^{-1/4}, \qquad \text{ or } \qquad \bar E_0 = \sqrt{1-\lambda_0^{-4}},
    \label{initial}
\end{equation}
where $\bar E_0 = \sqrt{\varepsilon/\mu_f}(V_0/H)$ is a non-dimensional measure of the voltage at this equilibrium.

\subsection{Exact bifurcation}
\label{Exact bifurcation}

To model the small-amplitude wrinkles appearing at the onset of lineari{\color{black}z}ed instability, we assume sinusoidal variations in the $x_1$-direction with wavelength $2\pi/k$, where $k$ is the {\color{black}wrinkling} wavenumber, and introduce the generalized, non-dimensional displacement-traction vector  $\bm{\eta}  (kx_2)= \left[ U_1, U_2, \Delta,  S_{21}, S_{22}, \Phi \right]^\text{T}$, where {\color{black}$U_1$, $U_2$ are the components of the incremental} mechanical displacement vector, $\Delta$ is a measure of the {\color{black}incremental} electric displacement {\color{black}in the $x_2$-direction}, {\color{black}$S_{21}$ and $S_{22}$} are the components of a measure of the incremental mechanical traction, and $\Phi$ is a measure of the incremental electric potential ({\color{black} all quantities depend} on the variable $kx_2$ only).

In the film and substrate, the incremental equations of equilibrium can be formulated as a first-order differential equation, $\bm{\eta}' = \rm{i} \bf N \bm{\eta}$,  where {\color{black}$\rm{i}=\sqrt{-1}$ is the imaginary unit, the prime denotes differentiation with respect to $kx_2$, and} $\bf N$ is the (constant) Stroh matrix, with components given explicitly in~\ref{AppeA}.
It is then straightforward to solve the boundary value problem (decay condition as $x_2 \to \infty$, continuity of $\bm \eta$ at the interface $x_2=0$, and {\color{black}the conditions of zero traction and a constant applied voltage on the top surface $x_2=-h$}).
As shown in~\ref{AppeA}, the exact bifurcation equation can be put {\color{black}into} the compact form
\begin{equation} \label{3.3.9}
    \det \left( \mathbf Z_f - r \mathbf Z_s \right) = 0,
\end{equation}
where $r = \mu_s/ \mu_f$ is the substrate-to-film \emph{stiffness ratio}, and $\mathbf Z_f$ and $\mathbf Z_s$ denote the impedance matrices of the film and substrate, respectively.
This equation depends only on the non-dimensional quantities $r$, $\lambda$, $kh$, and $\bar E_L = \sqrt{\varepsilon/\mu_f}(V/H)$.

We may then plot the dispersion curves (also referred to as bifurcation curves) for a given stiffness ratio: either the $\lambda-kh$ curves when we are interested in wrinkling instability under an increasing mechanical compression for a given voltage, or the $\bar E_L-kh$ curves when we focus on an increasing electric load for a given stretch.
Typically, these curves exhibit an extremum: a maximum $\lambda = \lambda_\text{cr}$ in the former case, a minimum $\bar E_L= \bar E_L^\text{cr}$ in the latter case (although not always), see Section \ref{section6}.
These extrema are the sought critical stretches and critical voltages of primary bifurcation.

\subsection{Approximate bifurcation and critical fields}

In the Results section, we show that the critical fields occur in the early part of the $kh$ span, typically when $kh < 2$.
Moreover, the critical value $(kh)_\text{cr}$ decreases as the dielectric film becomes stiffer than the substrate (i.e., $r$ is small).
It thus makes sense to seek Taylor series expansions of the bifurcation condition \eqref{3.3.9}. {\color{black}As detailed in~\ref{AppeA}, we expand the relationship $\bm \eta(-kh) = \exp (- \text i kh \mathbf N) \bm \eta(0)$} up to the sixth power in $kh$, apply the incremental boundary conditions, and observe that the resultant approximate expansion captures accurately the extrema corresponding to the critical wrinkling values in the early parts of the bifurcation curves.

Furthermore, under the assumption that the film is significantly stiffer than the substrate (i.e., $r$ is small, of order $(kh)^3$), we perform an asymptotic analysis based on the sixth-order approximate expansion to obtain explicit asymptotic expressions in $kh$ for the stretch $\lambda$ and the voltage $\bar E_{L}$. Then we can derive asymptotic expansions of the critical wavenumber $(kh)_\text{cr}$ and critical loads $\lambda_\text{cr}$ (stretch) and {\color{black}$\bar E_{L}^\text{cr}$} (voltage) explicitly by finding the first extremum of the $kh$-polynomial asymptotic expressions for the stretch $\lambda$ and voltage $\bar E_{L}$. Each of these critical quantities can be expressed in an $r^{1/3}$ power series, allowing us to extend the scaling laws of \cite{allen1969analysis} to electro-elasticity. {\color{black}The detailed derivation is presented in~\ref{AppeB}, with explicit asymptotic expansions provided in Section \ref{section3.3}.}

\subsection{Numerical post-buckling analysis}
\label{sec:numerics}

Complex nonlinear behaviors may be expected beyond the bifurcation, such as secondary bifurcations, period doubling, and self-contact folding \citep{brau2011multiple,fu2015asymptotic}.
Here, we rely on the finite element method to investigate the post-buckling behavior of the {\color{black}soft dielectric film-substrate system}.

We use a quasi-incompressible formulation of the problem to avoid element locking, with the following energy functionals for the substrate and the dielectric film, respectively,
\begin{equation}
    \label{eq:energy_functionals}
    \mathcal{E}_s[\vect{u}] = \int_{\mathcal{B}_s} \widehat W_s(\mathbf{F})\,\text{d}V,\qquad
    \mathcal{E}_f[\vect{u},\,\varphi] = \int_{\mathcal{B}_f} \widehat{W}_f(\mathbf{F},\,\vect{E}_L)\,\text{d}V,
\end{equation}
where $\vect{u}$ and $\varphi$ are the mechanical displacement {\color{black}vector} and the electric-potential field, $\mathbf{F}$ is the deformation gradient (two-point) {\color{black}tensor}, $\vect{E}_L = {\color{black}-} \Grad \varphi$ is the Lagrangian electric field {\color{black}vector}, $\widehat W_s$ and $\widehat W_f$ are the strain energy densities of the substrate and dielectric film, respectively, and $\mathcal{B}_s$ and $\mathcal{B}_f$ are the domains of the substrate and film.

For the nearly incompressible versions of the {\color{black}total} energy densities \eqref{eq:energies}, we take
\begin{equation}
    \widehat{W}_s = \tfrac{1}{2} \mu_s (I_1^* - 3) + \tfrac{1}{2}K_s (\ln J)^2,
    \qquad
    \widehat{W}_f = \tfrac{1}{2}\mu_f (I_1^* - 3) + \tfrac{1}{2}K_f (\ln J)^2 - \tfrac{1}{2}\varepsilon J\vect{E}:\vect{E},
\end{equation}
where $\vect{E} = \tens{F}^{-\text{T}}\vect{E}_L$ is the Eulerian electric field {\color{black}vector}, $J = \det(\tens{F})$, $I_1^* = J^{-2/3}\tr(\tens{F}^\text{T} \tens{F})$ and the $K_i$ {\color{black}($i=s,f$)} are the {\color{black}initial} bulk moduli (chosen {\color{black}to be} much larger than the $\mu_i$).
It can be shown that the stationary points of the total energy functional $\mathcal{E} = \mathcal{E}_f + \mathcal{E}_s$ correspond to equilibrium configurations of the system, see \cite{toupin1956elastic} and \cite{dorfmann2014nonlinear}. To approximate the fully incompressible case, we set $K_i = 500 \mu_i$ ($i = s, f$), so that the initial Poisson ratio is
\begin{equation}
    \nu_i = \frac{3K_i-2\mu_i}{2(3K_i+\mu_i)}\simeq 0.499, \qquad (i=s,f).
\end{equation}

The system is modeled by using a rectangular computational domain $[0,\,L]\times[{\color{black}D,\,-H}]$, where the depth $D$ {\color{black}of the substrate} is large compared to the film thickness $H$ (specifically, we set $D = 30 H$), and the length $L$ is chosen as half of the wavelength of the wrinkling pattern. On the left and right sides of the domain, we impose symmetry boundary conditions to mimic the infinite {\color{black}layered} half-space. We use a triangular structured mesh in the {\color{black}dielectric} film (with at least ten elements along the width), while the mesh is unstructured in the substrate, with a coarser mesh as we move away from the film-substrate interface.

To track the bifurcated branch, an arclength continuation algorithm is used, as described in \cite{Su_2023Tunable}, where either the stretch $\lambda$ or the non-dimensional voltage $\bar{E}_L$ are used as the control parameter. However, the handling of a control parameter that enters the boundary conditions is not straightforward using the arclength continuation algorithm. To avoid this issue, we split the displacement field additively as $\vect{u} = \vect{u}_h + \vect{u}_i$,
where $\vect{u}_h = (\lambda -1) X_1 \vect{\hat{I}}_1+ (\lambda^{-1} - 1) X_2\vect{\hat{I}}_2$ is the displacement field corresponding to the homogeneous deformation, and $\vect{u}_i$ is the inhomogeneous displacement field corresponding to the wrinkles, {\color{black}with $\vect{\hat{I}}_1$ and $\vect{\hat{I}}_2$ representing the unit basis vectors of the initial undeformed configuration}.
As $\vect{u}_h$ is known, we can solve the problem with respect to $\vect{u}_i$ to reconstruct the full displacement field. A similar splitting is performed for the electric potential, i.e., $\varphi = \varphi_h + \varphi_i$ with $\varphi_h = - V X_2$. The Dirichlet boundary conditions on $\vect{u}_i$ and $\varphi_i$ are homogeneous, {\color{black}facilitating a more straightforward} implementation of the arclength method (see also \cite{riccobelli2024elastic}).
A piecewise quadratic polynomial basis {\color{black}is employed} for the inhomogeneous displacement, {\color{black}while} a piecewise linear polynomial basis {\color{black}is used} for the electric potential. A small imperfection is applied to the top surface to trigger the instability.

We solve the problem using the finite element method implemented in \texttt{FEniCS} \citep{alnaes2015fenics}, which allows for automatic differentiation of the weak form and efficient assembly of the linear system. We obtain the solution by solving a Newton-Raphson problem, where the Jacobian matrix is computed using the automatic differentiation library \texttt{UFL}. For the continuation algorithm, we use the arclength method implemented in \texttt{BiFEniCS}.

\section{Results and discussion}
\label{section6}

\subsection{Critical bifurcation stretch for a prescribed electric field}
\label{section6.2.1}

First we consider the scenario where the applied voltage is fixed at its initial traction-free equilibrium value $V_0$, and the film-substrate system is deformed homogeneously under the action of a uniaxial stress along $x_1$.
Hence, the total Cauchy stresses $\sigma_f = \mu_f(\lambda^2 - \lambda^{-2}) - \varepsilon \lambda^2(V_0/H)^2$ and $\sigma_s = \mu_s(\lambda^2 - \lambda^{-2})$ are applied to the dielectric film and the substrate, respectively.

\begin{figure}[t!]
    \centering
    \setlength{\abovecaptionskip}{5pt}
    \includegraphics[width=0.98\textwidth]{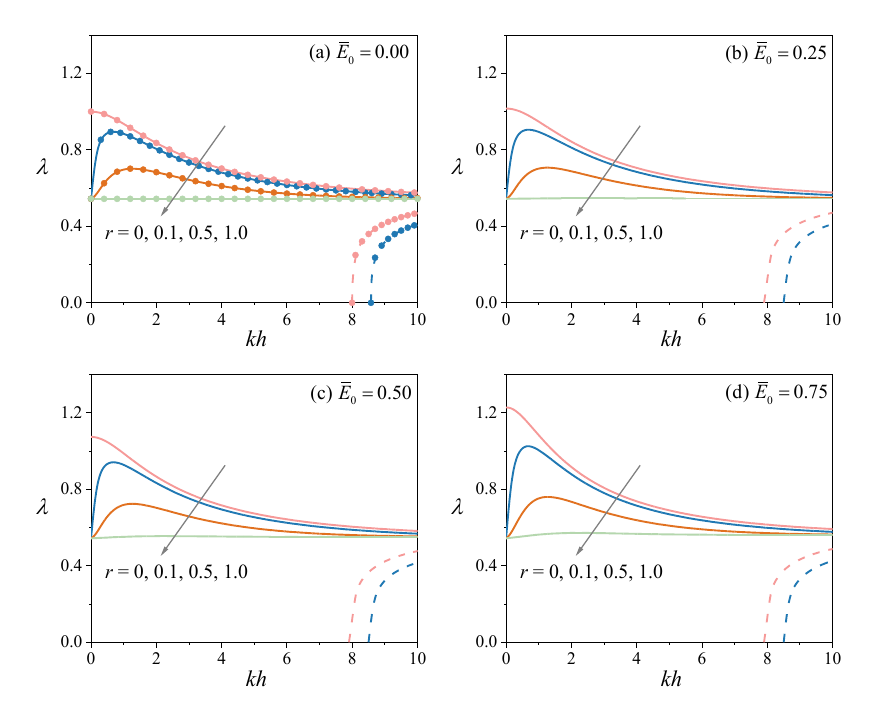}
    \caption{Bifurcation curves of stretch $\lambda$ as a function of $kh$ with different substrate-to-film shear modulus ratios $r = \mu_s/\mu_f = 0, 0.1, 0.5, 1.0$ under four non-dimensional voltages applied to the soft dielectric film, {\color{black}subjected to a plane-strain load}: (a) ${\bar E}_0=0$; (b) ${\bar E}_0=0.25$; (c) ${\bar E}_0=0.5$; (d) ${\bar E}_0=0.75$.
            {\color{black}Solid curves: antisymmetric-dominated modes; Dashed curves: symmetric-dominated modes.
                The onset of instability occurs at the maximum of the bifurcation curve, always corresponding to an antisymmetric-dominated mode. In Fig.~\ref{Fig6-3}(a), solid lines show the proposed surface impedance results, and symbols indicate the displacement solution predictions \citep{cai1999imperfection}.}}
    \label{Fig6-3}
\end{figure}

Fig.~\ref{Fig6-3} shows the bifurcation curves of stretch $\lambda$ against wavenumber $kh$ for different shear modulus ratios $r = \mu_s/\mu_f$ under four given non-dimensional voltages ${\bar E}_0=0, 0.25, 0.5, 0.75$. {\color{black} When ${\bar E}_0=0$, we have excellent agreement with \cite{cai1999imperfection}, who used the classical displacement solution method for a coated hyperelastic half-space, confirming the effectiveness of the surface impedance method, see Fig.~\ref{Fig6-3}(a).}
All bifurcation curves, except for the $r=0$ one, start at $\lambda=0.5437$, the Biot critical stretch of instability for an elastic neo-Hookean half-space, which corresponds to $kh=0$ (vanishing film thickness).
When $r=0$, there is no substrate; then, in the $kh \to 0$ limit, the dielectric film is infinitesimally thin and {\color{black}wrinkles immediately once} $\sigma_f$ is applied.
Hence, the limit is found by solving $\sigma_f=0$, which according to Eq.~\eqref{initial}, gives the limits $\lambda = 1.000, 1.016, 1.075, 1.230$ {\color{black}for} $\bar E_0=0, 0.25, 0.5, 0.75$, respectively.
Conversely, in the $kh \to \infty$ limit, the {\color{black}soft dielectric} film becomes a semi-infinite ideal dielectric, and all bifurcation curves tend to the root of
\begin{equation}
    \lambda^6 + \lambda^4 + 3\lambda^2 - 1 = \lambda^4(1+\lambda^2)\bar E_0^2,
    \label{surf-inst-plane}
\end{equation}
{\color{black}which corresponds to} the surface instability criterion in plane strain (see \ref{AppeA} for more general formulas and details).
It {\color{black}yields} $\lambda \to 0.5437, 0.5454, 0.5508, 0.5607$ when $\bar E_0=0, 0.25, 0.5, 0.75$, respectively.

\begin{table}[t!]
    \centering
    \begin{tabular}{lllllll}
        \toprule
                                         & Value for       & Value for         & Value for         & Value for          & Value for         & Value for         \\
                                         & $\bar{E}_0 = 0$ & $\bar{E}_0 = 0.3$ & $\bar{E}_0 = 0.6$ & $\bar{E}_0 = 0.75$ & $\bar{E}_0 = 0.9$ & $\bar{E}_0 = 1.0$ \\
        \midrule
        \multicolumn{7}{l}{\textbf{Critical values when $r=1/5$}}                                                                                               \\ %
        $\lambda_\text{cr}$              & 0.8354          & 0.8481            & 0.8933            & 0.9372             & 1.0119            & 1.1005            \\
        $\lambda^\text{num}_\text{cr}$   & 0.8353          & 0.8481            & 0.8936            & 0.9378             & 1.0134            & 1.1085            \\
        $\lambda^\text{asymp}_\text{cr}$ & 0.8569          & 0.8756            & 0.9605            & 1.1097             & 2.3088            & $--$              \\
        $(kh)_\text{cr}$                 & 0.88            & 0.88              & 0.88              & 0.88               & 0.86              & 0.86              \\
        $(kh)^\text{asymp}_\text{cr}$    & 0.87            & 0.88              & 0.89              & 0.91               & 0.99              & $--$              \\
        super/subcritical                & super-          & super-            & super-            & super-             & super-            & sub-              \\
        \addlinespace %
        \multicolumn{7}{l}{\textbf{Critical values when $r=1/30$}}                                                                                              \\ %
        $\lambda_\text{cr}$              & 0.9482          & 0.9674            & 1.0401            & 1.1190             & 1.2822            & 1.5874            \\
        $\lambda^\text{num}_\text{cr}$   & 0.9486          & 0.9680            & 1.0408            & 1.1200             & 1.2842            & 1.6013            \\
        $\lambda^\text{asymp}_\text{cr}$ & 0.9493          & 0.9689            & 1.0435            & 1.1277             & 1.3548            & $--$              \\
        $(kh)_\text{cr}$                 & 0.46            & 0.46              & 0.46              & 0.46               & 0.44              & 0.44              \\
        $(kh)^\text{asymp}_\text{cr}$    & 0.47            & 0.47              & 0.46              & 0.46               & 0.45              & $--$              \\
        super/subcritical                & super-          & super-            & super-            & super-             & super-            & sub-              \\
        \bottomrule
    \end{tabular}
    \caption{{\color{black}Critical stretch and wavenumber values, together with the classification of bifurcation type (supercritical or subcritical), for different electric loadings $\bar{E}_0$ and shear modulus ratios $r$. The notation ``$--$'' indicates cases where the asymptotic solution is inapplicable. Superscripts ``num'' and ``asymp'' represent the critical values calculated by finite element numerical simulations and the asymptotic expansion expressions \eqref{4.2.5}-\eqref{4.2.4}, respectively.}}
    \label{tab:combined_critical_values}
\end{table}

Between these two limits, the {\color{black}bifurcation} curve goes through a maximum, which determines the critical stretch $\lambda_\text{cr}$ and critical wavenumber $(kh)_\text{cr}$.
We collected these critical values for six given voltages and two stiffness ratios in Table \ref{tab:combined_critical_values}.
When there is no applied voltage (${\bar E}_0= 0$), the values of the critical compressive strain are $\epsilon_\text{w} = 1 - \lambda_\text{cr} = 0.165$ and $0.052$ for $r=1/5$ and $1/30$, respectively, in reasonable agreement with \cite{cao2012Wrinkling}.
We observe {\color{black}from Fig.~\ref{Fig6-3} and Table \ref{tab:combined_critical_values} that, for a fixed modulus ratio $r$, the critical stretch $\lambda_\text{cr}$ monotonically increases with increasing applied voltage ${\bar E}_0$, and that, for a prescribed voltage, $\lambda_\text{cr}$ also increases as $r$ decreases. While the applied voltage exerts only a negligible influence on the critical wavenumber $(kh)_{\rm cr}$, the latter progressively decreases with decreasing modulus ratio.}

\begin{figure}[t!]
    \centering
    \setlength{\abovecaptionskip}{5pt}
    \includegraphics[width=0.98\textwidth]{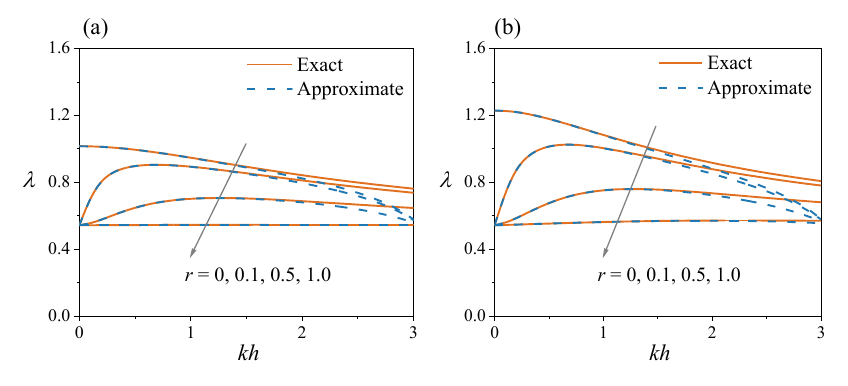}
    \caption{Exact and approximate (sixth-order) bifurcation curves of stretch $\lambda$ as functions of $kh$ {\color{black}for different shear modulus ratios $r=\mu_s/\mu_f$ under two non-dimensional voltages, (a) ${\bar E}_0=0.25$ and (b) ${\bar E}_0=0.75$, showing that the approximations capture the critical points accurately, thus enabling asymptotic expansions of $\lambda_\text{cr}$ and $(kh)_\text{cr}$ in terms of $r$ and $\bar E_0$.}}
    \label{Fig6-4}
\end{figure}

{\color{black}To further elucidate the influence of the applied voltage $\bar E_0$ on $\lambda_\text{cr}$ and $(kh)_\text{cr}$, we seek their asymptotic expansions in terms of $r$ and $\bar E_0$.
But prior to that, Fig.~\ref{Fig6-4} shows that the bifurcation curves of stretch $\lambda$ versus $kh$ predicted by the sixth-order Taylor approximate solution \eqref{4.2.1} agrees remarkably well with the exact bifurcation equation \eqref{3.3.9} over a broad range of $kh$, thereby validating its accuracy in capturing the extrema of the bifurcation curves and providing a solid foundation for an asymptotic analysis of the critical parameters $\lambda_\text{cr}$ and $(kh)_\text{cr}$.}

Then we use {\color{black}the sixth-order approximation to perform the asymptotic analysis (see \ref{AppeB} and Section \ref{section3.3}) and derive the explicit leading-order correction} for the relative extension of wrinkling instability due to electro-mechanical loading, as
\begin{equation}
    \epsilon_\text{w} = \left| \frac{\lambda_\text{cr} - \lambda_0}{\lambda_0}\right | = \frac{1}{4}\frac{1}{1 - \bar E_0^2}\left(\frac{1+\sqrt{1 - \bar E_0^2}}{2} \, 3 \frac{\mu_s}{\mu_f}\right)^{2/3},
    \label{extension}
\end{equation}
{\color{black}where we recall that $\bar E_0 \leq 1$.
This expression recovers the classical $2/3$ power scaling law with respect to the modulus ratio $r={\mu_s}/{\mu_f}$ for the purely elastic case, as originally reported by \cite{allen1969analysis}, while additionally revealing that the multiplicative factor associated with $\bar E_0$ is a monotonically increasing function of the applied voltage. This indicates that a higher initial voltage permits a larger relative compressive strain, quantified as $\lambda_0 - \lambda_\text{cr}$, to trigger instability. In other words,  once the initial traction-free voltage $\bar E_0$ is applied, the system can sustain greater homogeneous deformation before the onset of instability, thereby postponing the emergence of wrinkles.
Note that using Eq.~\eqref{initial}, the result can equivalently be expressed} in terms of the initial {\color{black}traction-free} pre-stretch $\lambda_0$ as
\begin{equation}
    \epsilon_\text{w} = \frac{\lambda_0^4}{4}\left(\frac{1 + \lambda_0^{-2}}{2} \, 3 \frac{\mu_s}{\mu_f}\right)^{2/3}.
\end{equation}

{\color{black}Similarly, the scaling relation governing the leading-order correction to the critical wavenumber is derived as}
\begin{equation}
    (kh)_\text{cr} = \left(\frac{1+\sqrt{1-E_0^2}}{2}\, 3 \frac{\mu_s}{\mu_f}\right)^{1/3} = \left(\frac{1+\lambda_0^{-2}}{2}\, 3 \frac{\mu_s}{\mu_f}\right)^{1/3},
\end{equation}
see \ref{AppeB} and Section \ref{section3.3}, which provides details and further approximations and expansions.

\subsection{Critical bifurcation electric voltage for a fixed pre-stretch}
\label{section6.2.2}

Conversely, we may hold the pre-stretch at an initial fixed value $\lambda$ and observe wrinkling as the applied voltage $V$ changes.
In that scenario, the pre-load is achieved by applying the uniaxial Cauchy stresses $\sigma_f = \mu_f(\lambda^2 - \lambda^{-2}) - \varepsilon \lambda^2(V/H)^2$ in the dielectric film and $\sigma_s = \mu_s(\lambda^2 - \lambda^{-2})$ in the substrate.

\begin{figure}[h]
    \centering
    \setlength{\abovecaptionskip}{5pt}
    \includegraphics[width=1.0\textwidth]{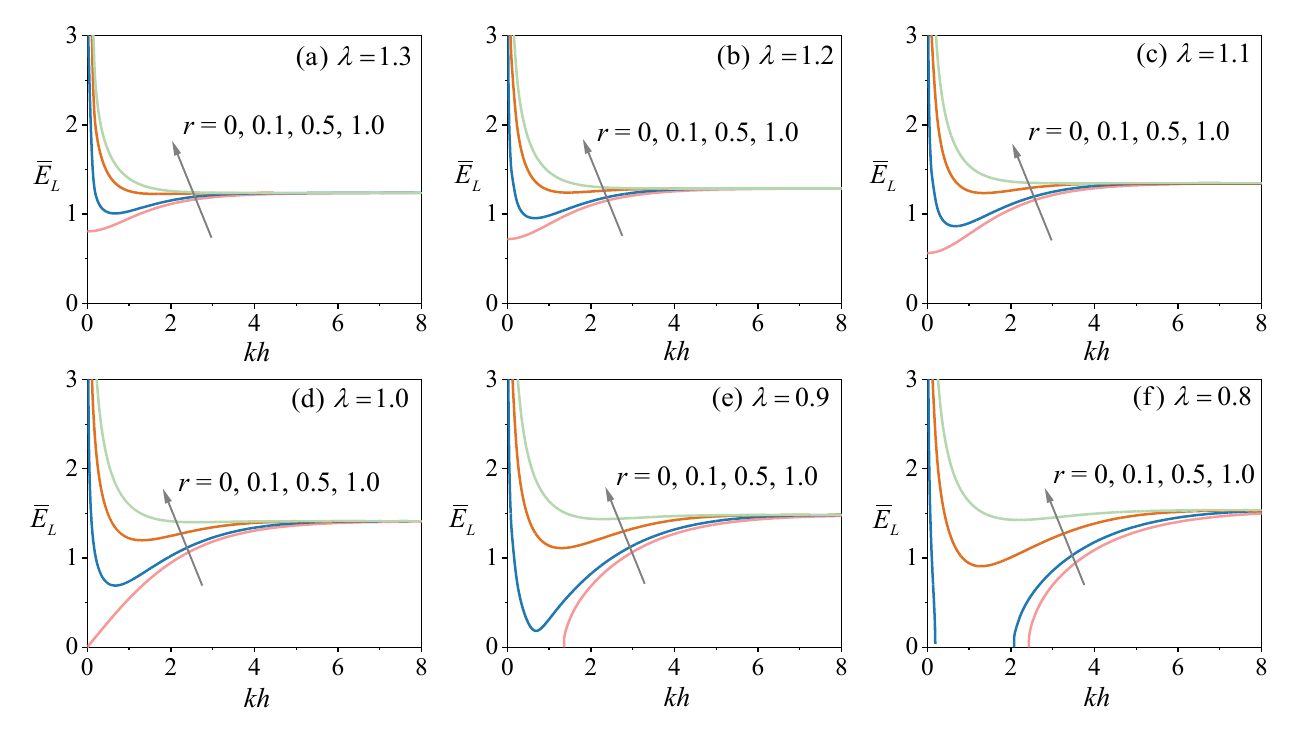}
    \caption{Bifurcation curves of the non-dimensional voltage ${\bar E}_L$ as a function of $kh$ {\color{black}for different stiffness ratios $r=0,0.1,0.5,1.0$ and six fixed pre-stretches: (a-c) extensile stretches $\lambda=1.3, 1.2, 1.1$; (d-f) contractile stretches $\lambda=1.0, 0.9, 0.8$.} The minima correspond to the critical voltage $\bar E_{L}^{\text{cr}}$ and critical wavenumber $(kh)^\text{cr}$. For a sufficiently contractile stretch ($\lambda< \lambda_{\rm c}^0$) and a sufficiently stiff film ($r<r_{\rm c}^0$), the system wrinkles before the application of voltage, as shown in Fig.~\ref{Fig6-5}(f) for $r = 0.1$, for example.}
    \label{Fig6-5}
\end{figure}

{\color{black}The bifurcation curves of the non-dimensional voltage $\bar E_L = \sqrt{\varepsilon/\mu_f}(V/H)$ as functions of the wavenumber $kh$ are presented in Fig.~\ref{Fig6-5} for various shear modulus ratios $r = \mu_s/\mu_f$ under six prescribed pre-stretch values $\lambda = 1.3, 1.2, 1.1, 1.0, 0.9, 0.8$. The results reveal that the bifurcation curves of $\bar E_L$ with respect to $kh$ generally exhibit non-monotonic behavior, except in the special case of $r = 0$, where the curve increases monotonically.}
However, all bifurcation curves exhibit a minimum, corresponding to the critical voltage $\bar E_{L}^{\text{cr}}$ of primary interest. Moreover, Fig.~\ref{Fig6-5} reveals that both decreasing pre-stretch $\lambda$ and reducing modulus ratio $r$ lead to a progressive decline in the critical voltage $\bar E_{L}^{\text{cr}}$, thereby indicating an increased susceptibility of the system to wrinkling instability.
In particular, if the pre-stretch is sufficiently contractile ($\lambda< 1$) and the film is sufficently stiff ($r$ small), we expect that wrinkling may occur for small values of the voltage loading, which is confirmed by the trend in Figs.~\ref{Fig6-5}(e) and \ref{Fig6-5}(f).

\begin{figure}[t!]
    \centering
    \setlength{\abovecaptionskip}{5pt}
    \includegraphics[width=0.98\textwidth]{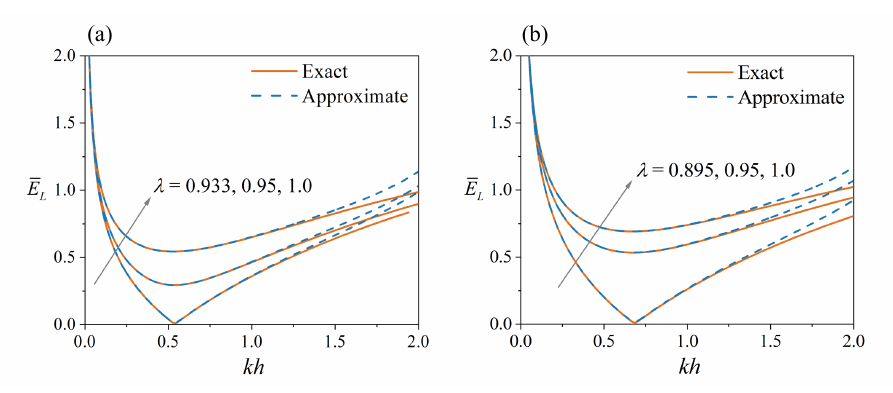}
    \caption{{\color{black}Exact and approximate (sixth-order) bifurcation curves of voltage $\bar E_L$ as functions of $kh$ for different pre-stretches $\lambda$ and two shear modulus ratios, (a) $r=0.05$ and (b) $r=0.1$, demonstrating that the approximations accurately capture the critical points, thus enabling asymptotic expansions of $\bar E_L^\text{cr}$ and $(kh)^\text{cr}$ in terms of $r$ and $\lambda$. For a pre-stretch $\lambda$ marginally exceeding $\lambda_{\rm c}^0=0.933$ (a) and $0.895$ (b), even a small applied voltage is sufficient to induce instability.}}
    \label{Fig7}
\end{figure}

{\color{black}To investigate the effect of the applied pre-stretch $\lambda$ on the critical voltage $\bar E_L^\text{cr}$ and the critical wavenumber $(kh)^\text{cr}$, we seek their asymptotic representations with respect to $r$ and $\lambda$. Again, the sixth-order Taylor expansion leads to an excellent agreement with the exact bifurcation condition across a wide interval of $kh$, establishing a rigorous basis for the asymptotic characterization of the critical quantities $\bar E_L^\text{cr}$ and $(kh)^\text{cr}$, see Fig.~\ref{Fig7}.}

{\color{black}Moreover, Fig.~\ref{Fig7} reveals that, for a prescribed modulus ratio $r$, a decrease in the applied pre-stretch $\lambda$ leads to a gradual reduction in the corresponding critical voltage $\bar E_{L}^{\text{cr}}$, a trend consistent with the behavior previously identified in Fig.~\ref{Fig6-5}. Particularly,} the critical voltage $\bar E_{L}^{\text{cr}}$ may eventually reach zero for $r=0.05$, $\lambda \simeq 0.933$, and $r=0.1$, $\lambda \simeq 0.895$, at which point we recover the critical values for the stretch and wavenumber of a purely elastic film-substrate system \citep{cai1999imperfection,cao2012Wrinkling}.
{\color{black}When $r$ or $\lambda$ is further reduced,} the incremental analysis breaks down, and the negative value of the minimum has no physical meaning (e.g., the case $\lambda = 0.8$ and $r=0.1$ in Fig.~\ref{Fig6-5}(f)).

\begin{figure}[t!]
    \centering
    \setlength{\abovecaptionskip}{5pt}
    \includegraphics[width=0.55\textwidth]{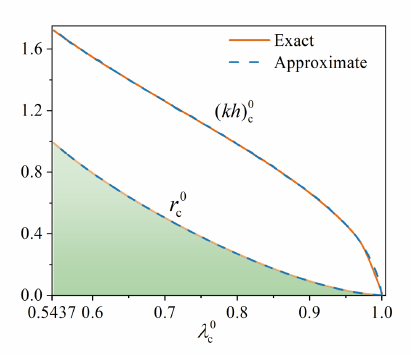}
    \caption{Lower curve: {\color{black}$\lambda_{\rm c}^0 - r_{\rm c}^0$ critical curve corresponding to a vanishing critical voltage (i.e., the purely elastic limit),} below which the film-substrate system wrinkles mechanically, prior to the application of a voltage. Upper curve: corresponding {\color{black}critical} wavenumber $(kh)_{\rm c}^0$. }
    \label{Fig6}
\end{figure}

Fig.~\ref{Fig6} shows the domain in the $\lambda-r$ plane where the soft dielectric film-substrate system can be expected to wrinkle under an applied voltage.
    {\color{black}The demarcation curve, referred to as the $\lambda_{\rm c}^0 - r_{\rm c}^0$ critical curve, corresponds to a vanishing critical electric field (i.e., the purely elastic limit). This critical curve} is found by solving the exact or the sixth-order approximate bifurcation condition {\color{black}when $\bar E_{L}^{\rm cr} = 0$}, and behaves asymptotically as
\begin{equation}
    {\color{black}\lambda_{\rm c}^0 = 1 - \tfrac{1}{4}(3r_{\rm c}^0)^{2/3},}
    \label{hyper}
\end{equation}
according to Eq.~\eqref{extension} written at $\bar E_0=0$ and $\lambda_0=1$, in agreement {\color{black}with} \cite{cai1999imperfection}.
In Fig.~\ref{Fig6}, we denote by ($\lambda_{\rm c}^0,r_{\rm c}^0$) the coordinates of points on that critical curve,  {\color{black}with $(kh)_{\rm c}^0$ representing the corresponding critical wavenumber.
        Parameter combinations of pre-stretch and stiffness ratio underneath that critical curve lead to wrinkling prior to the application of any voltage, whereas those above the critical curve require a finite applied voltage to trigger wrinkling instability.}

We also {\color{black}employ the sixth-order approximation and perform the asymptotic analysis (see \ref{AppeB} and Section \ref{section3.3}) to} obtain an explicit expression for the leading-order correction to the squared critical voltage in terms of $r$ and $\lambda$ as,
\begin{equation}
    (\bar E_{L}^{\text{cr}})^2 =  1 - \lambda^{-4} + \left(\frac{1 + \lambda^{-2}}{2} \, 3 r \right)^{2/3},
\end{equation}
provided the values of ($\lambda, r$) are not in the shaded area of Fig.~\ref{Fig6}.
This is equivalent to $(\bar E_{L}^{\text{cr}})^2 \ge 0$, or, by expansion for small $r$, $\lambda \ge 1 - (1/4)(3r)^{2/3}$, in agreement with Eq.~\eqref{hyper}.

We collected the critical values of voltage and wavenumber for six given pre-stretches and two stiffness ratios in Table \ref{tab:voltage_critical_values}.
{\color{black}It shows that a reduction in the pre-stretch $\lambda$ or a decrease in the modulus ratio $r$ diminishes the critical voltage $\bar E_{L}^{\text{cr}}$, thereby indicating an enhanced propensity of the soft dielectric film-substrate system to undergo wrinkling instability, as mentioned earlier.
In addition,} under pre-compression ($\lambda<1$), only a tiny voltage is required to trigger wrinkling when the dielectric film is much stiffer than the substrate. We also see that an applied (albeit larger) voltage can render the system unstable when it is pre-elongated ($\lambda >1$), because the film tends to expand in its plane under the applied voltage, which is prevented when $\lambda$ is fixed, eventually leading to wrinkles.

\begin{table}[t!]
    \centering
    \begin{tabular}{lllllll}
        \toprule
                                          & Value for        & Value for       & Value for        & Value for       & Value for       & Value for       \\
                                          & $\lambda = 0.85$ & $\lambda = 0.9$ & $\lambda = 0.95$ & $\lambda = 1.0$ & $\lambda = 1.1$ & $\lambda = 1.2$ \\
        \midrule
        \multicolumn{7}{l}{\textbf{Critical values when $r=1/5$}}                                                                                       \\ %
        $\bar E_{L}^{\text{cr}}$          & 0.3207           & 0.6280          & 0.7825           & 0.8814          & 0.9996          & 1.0535          \\
        $\bar E_\text{num}^{\text{cr}}$   & 0.3214           & 0.6278          & 0.7819           & 0.8808          & 0.9913          & 1.0038          \\
        $\bar E_\text{asymp}^{\text{cr}}$ & 0.3962           & 0.6605          & 0.8034           & 0.8967          & 1.0098          & 1.0724          \\
        $(kh)^\text{cr}$                  & 0.88             & 0.87            & 0.87             & 0.86            & 0.86            & 0.86            \\
        $(kh)^\text{cr}_\text{asymp}$     & 0.82             & 0.82            & 0.82             & 0.82            & 0.82            & 0.83            \\
        super/subcritical                 & super-           & super-          & super-           & super-          & sub-            & sub-            \\
        \addlinespace %
        \multicolumn{7}{l}{\textbf{Critical values when $r=1/30$}}                                                                                      \\ %
        $\bar E_L^\text{cr}$              & $--$             & $--$            & 0.0937           & 0.4729          & 0.7213          & 0.8407          \\
        $\bar E_\text{num}^{\text{cr}}$   & $--$             & $--$            & 0.0949           & 0.4723          & 0.7210          & 0.8407          \\
        $\bar E_\text{asymp}^{\text{cr}}$ & $--$             & $--$            & 0.0985           & 0.4732          & 0.7216          & 0.8409          \\
        $(kh)^\text{cr}$                  & $--$             & $--$            & 0.47             & 0.46            & 0.45            & 0.45            \\
        $(kh)^\text{cr}_\text{asymp}$     & $--$             & $--$            & 0.47             & 0.46            & 0.45            & 0.44            \\
        super/subcritical                 & $--$             & $--$            & super-           & super-          & super-          & super-          \\
        \bottomrule
    \end{tabular}
    \caption{{\color{black}Critical voltage and wavenumber values, together with the classification of bifurcation type (supercritical or subcritical), for different mechanical loadings $\lambda$ and shear modulus ratios $r$. The notation ``$--$'' indicates cases where the wrinkling has occurred prior to the application of the voltage. Subscripts ``num'' and ``asymp'' represent the critical values calculated by finite element numerical simulations and the asymptotic expansion expressions \eqref{4.2.7}-\eqref{19}, respectively.}}
    \label{tab:voltage_critical_values}
\end{table}

\subsection{Asymptotic expansions for high-contrast stiffness ratios} \label{section3.3}

The numerical root-finding procedure for the exact bifurcation equation \eqref{3.3.9} {\color{black}incurs} a heavy computational cost, such that asymptotic expansions may provide a much-needed rapid alternative way to find the critical values of stretch and voltage.

For small $r$ and $kh$, and under the assumption that $r$ is of order $(kh)^3$, we show in \ref{AppeB} how a {\color{black}fourth-order series expansion} can be derived for the stretch {\color{black}through asymptotic analysis},
\begin{multline} \label{4.2.2}
    \frac{\lambda}{\lambda_0} = 1   - \frac{1}{4}\lambda_0^2(1+\lambda_0^2) \left(\frac{r}{kh}\right)  - \frac{1}{12}\lambda_0^4 (kh)^2 + \frac{1}{4}(\lambda_0^2-1)r \\[4pt]
    + \frac{1}{48}(\lambda_0^2-1)(3 + 7\lambda_0^2  +8 \lambda_0^4 + 5 \lambda_0^6) r(kh)
    + \frac{1}{1440}\lambda_0^4(2+37\lambda_0^4)(kh)^4\\[4pt]
    + \frac{1}{32} \left(2 - 4\lambda_0^2 +3\lambda_0^4 + 6\lambda_0^6 + 5 \lambda_0^8\right)\left(\frac{r}{kh}\right)^2,
\end{multline}
where $\lambda_0  = (1 - \bar E_0^2)^{-1/4}$ and we neglect terms of order $(kh)^6$ and higher. Then, by differentiating Eq.~\eqref{4.2.2} with respect to $kh$, we find where the stretch is (locally) maximized in the bifurcation curve.
    {\color{black}Disregarding contributions of order $r^{4/3}$ and higher, the expression for the critical wavenumber is obtained as}
\begin{equation} \label{4.2.5}
    (kh)_\text{cr} = \left(\frac{1 + \lambda_0^{-2}}{2} 3r \right)^{1/3} + \frac{12\lambda_0^{10}+54\lambda_0^8+19\lambda_0^6+19\lambda_0^4-53\lambda_0^2-15}
    {120\lambda_0^4(1+\lambda_0^2)} r,
\end{equation}
and neglecting terms of order $r^2$ and higher, the critical stretch is given by
\begin{multline} \label{4.2.4}
    \frac{\lambda_\text{cr}}{\lambda_0} = 1 - \frac{1}{4} \lambda_0^4\left(\frac{1 + \lambda_0^{-2}}{2} 3r \right)^{2/3} + \frac{1}{4}(\lambda_0^2-1)r
    \\[4pt]
    +  \frac{1}{1920}\left(237\lambda_0^{10} + 354\lambda_0^8 + 139\lambda_0^6 - 176\lambda_0^4 - 98\lambda_0^2 - 60\right)\left(\frac{2}{\lambda_0(1 + \lambda_0^2)}\right)^{2/3} r(3 r)^{1/3}.
\end{multline}
{\color{black}Given that $\lambda_0 = (1-\bar E_0^2)^{-1/4} \geq 0$, the validity of Eqs.~\eqref{4.2.2}-\eqref{4.2.4} is restricted to the regime $\bar E_0 \leq 1$. This regime, $\bar E_0 \leq 1$, also corresponds to the parameter range where we expect wrinkles to arise under a small electric voltage.}

In the absence of {\color{black}an} applied voltage ($\bar E_0 = 0$, $\lambda_0=1$), we recover the asymptotic formulas of the hyperelastic film-substrate system \citep{cai1999imperfection, Alawiye_2019Revisiting} for the stretch,
\begin{equation}\label{B6}
    \lambda  = 1 - \frac{1}{2}\left( {\frac{r}{{kh}}} \right) - \frac{1}{{12}}{\left( {kh} \right)^2} + \frac{3}{8}{\left( {\frac{r}{{kh}}} \right)^2} + \frac{{13}}{{480}}{\left( {kh} \right)^4},
\end{equation}
and for the critical stretch and critical wavenumber,
\begin{equation}
    \lambda_{\mathrm{cr}}
    = 1 - \frac{1}{4}(3r)^{2/3}
    + \frac{33}{160}\,r\,(3r)^{1/3},\qquad
    (kh)_{\mathrm{cr}}
    = (3r)^{1/3}
    + \frac{3}{20}\,r.
    \label{hypercrit}
\end{equation}

Similarly, {\color{black}we obtain the asymptotic expansion of the squared non-dimensional voltage in the form}
\begin{multline} \label{4.2.3}
    \bar E_L^2  =1 - \lambda^{-4}
    + \frac{1}{3}(kh)^2
    + (1 + \lambda^{-2})\left(\frac{r}{kh} \right)
    +(\lambda^{-4}-\lambda^{-2})r
    \\[4pt]
    - \frac{1}{4}(1 - \lambda^{-2})^2\left(\frac{r}{kh}\right)^2 + \frac{1}{12}(1 + \lambda^{-2})(1 + 3\lambda^{-2})\, r (kh)
    - \frac{1}{180}(1 + 6\lambda^4)(kh)^4.
\end{multline}
In the absence of voltage, $\bar E_L=0$ and we recover Eq.~\eqref{B6}.
{\color{black}By setting the derivative of Eq.~\eqref{4.2.3} with respect to $kh$ to zero, we determine the location of the (local) minimum.
Neglecting terms of order $r^2$ and higher, the asymptotic expression for the squared critical voltage is obtained as}
\begin{multline} \label{4.2.7}
    (\bar E_{L}^\text{cr})^2 =  1 - \lambda^{-4} + \left(\frac{1 + \lambda^{-2}}{2} \, 3 r \right)^{2/3}
    -(\lambda^{-2} - \lambda^{-4})r \\[4pt]
    - \frac{6\lambda^{10} + 12\lambda^{8} + 17\lambda^{6} - 88\lambda^{4} - 49\lambda^{2} - 30}
    {40 \cdot 2^{1/3}\cdot 3^{2/3}\,\lambda^{14/3}\,(\lambda^{2}+1)^{2/3}}{r^{4/3}},
\end{multline}
an expression that is valid provided $\lambda$ is greater than the right-hand side of Eq.~\eqref{hypercrit}$_1$.
Again, we may {\color{black}verify} that when $\bar E_L^\text{cr}=0$, Eq.~\eqref{4.2.7} is consistent with Eq.~\eqref{hypercrit}$_1$.
The asymptotic expansion for the corresponding critical wavenumber is
\begin{equation} \label{19}
    (kh)^\text{cr} = \left(\frac{1 + \lambda^{-2}}{2} 3r \right)^{1/3} + \frac{12\lambda^{10}+24\lambda^8 - 11\lambda^6 + 19\lambda^4 - 53\lambda^2-15}
    {120\lambda^4(\lambda^2+1)} r.
\end{equation}

\begin{figure}[t!]
    \centering
    \setlength{\abovecaptionskip}{5pt}
    \includegraphics[width=0.95\textwidth]{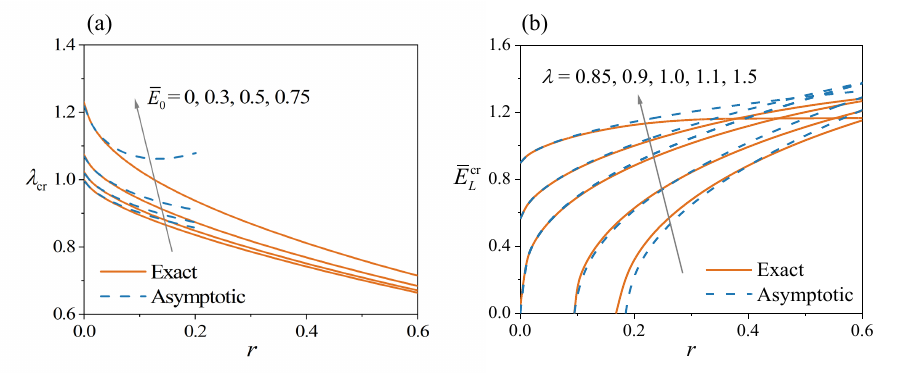}
    \caption{{\color{black}Variations in critical fields with the shear modulus ratio $r=\mu_s/\mu_f$: (a) critical stretch $\lambda_\text{cr}$ for different applied voltages $\bar E_0$; (b) critical voltage $\bar E_L^\text{cr}$ for various pre-stretches $\lambda$. Solid curves: exact solutions given by Eq.~\eqref{3.3.9}; Dashed curves: asymptotic expansions provided by Eqs.~\eqref{4.2.4} and \eqref{4.2.7}.} Asymptotic expansions of the critical fields provide a fast alternative to solving the exact bifurcation criterion when the soft dielectric film is much stiffer than the substrate ($r$ small).}
    \label{Fig-asymp}
\end{figure}

The validity of the asymptotic expansions provided by Eqs.~\eqref{4.2.4} and \eqref{4.2.7} is illustrated in Fig.~\ref{Fig-asymp}, exhibiting excellent agreement for small values of $r$ and reasonable accuracy for moderate $r$.

\subsection{Post-buckling analysis by the finite element method}
\label{section6.6}

\begin{figure}[t!]
    \centering
    \begin{minipage}{0.6\textwidth}\small(a)\\\includegraphics[width=\textwidth]{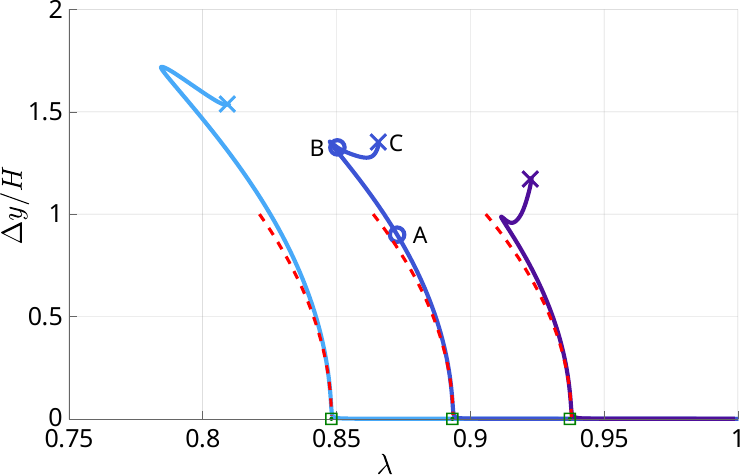}
    \end{minipage}

    \vspace{.5em}
    \begin{minipage}{\textwidth}\small(b)\\
        \includegraphics[width=\textwidth]{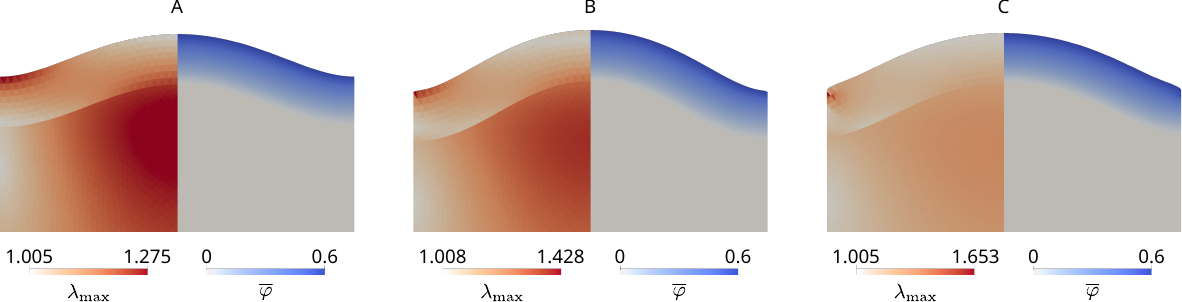}
    \end{minipage}
    \caption{(a) Plot of the {\color{black}non-dimensional} amplitude of the wrinkling of the free surface, $\Delta y/H$, against the stretch $\lambda$ for $r = 1/5$ and $\bar{E}_0 = 0.3$ (light blue line), $0.6$ (blue line), and $0.75$ (purple line). The green square denotes the marginal stability threshold obtained from the {\color{black}linearized} stability analysis, while the cross indicates the onset of self-contact of the free surface. Letters mark the positions on the bifurcation diagram corresponding to the configurations shown below.
        The dashed lines show the best fit of the finite element data close to the bifurcation point with the function ${\hat A} \sqrt{\lambda^\text{num}_\text{cr}-\lambda}$, with ${\hat A},\,\lambda^\text{num}_\text{cr}>0$; see the main text for details. Here, the fitted parameter ${\hat A} = 6.0952,\, 5.7925,\, 5.5954$ and $\lambda^\text{num}_\text{cr}=0.8481,\,0.8936,\,0.9378$ for $\bar{E}_0=0.3,\,0.6,\,0.75$, respectively.
        (b) Deformed configurations corresponding to the points $A,\,B,\,C$ indicated in the bifurcation diagram, where the maximum principal stretch $\lambda_{\text{max}}$ (i.e., the square root of the maximum eigenvalue of $\mathbf{F}^\text{T}\mathbf{F}$ for each point of the domain) and the {\color{black}non-dimensional electric potential field}, $\bar{\varphi} = \sqrt{\varepsilon/\mu_f}(\varphi/H)$, are reported. A progressive strain localization is observed in the furrows of the wrinkling pattern beyond the turning point $B$, eventually leading to crease formation in the point $C$.}
    \label{fig:bif_r_1_5}
\end{figure}

We now turn to post-buckling analysis.
First we plot the {\color{black}non-dimensional} amplitude of the free surface wrinkles, $\Delta y/H$, as a function of the stretch $\lambda$ for $r = 1/5$ and various {\color{black}non-dimensional} electric voltages $\bar{E}_0 = 0.3, 0.6, 0.75$, see Fig.~\ref{fig:bif_r_1_5}(a).
Here, $\Delta y$ denotes the difference between the maximum and minimum vertical positions of points on the free surface. The resulting diagrams are reminiscent of supercritical pitchfork bifurcations.

To verify this hypothesis, we fit the following function
\begin{equation}
    \label{eq:fitting}
    \frac{\Delta y}{H} = \hat A\sqrt{|\lambda^\text{num}_\text{cr} - \lambda|}
\end{equation}
to the finite element numerical data near the bifurcation point. Such a {\color{black}parabolic} function represents the amplitude of the wrinkling pattern for a pitchfork bifurcation in the weakly nonlinear regime. From the numerical simulations, we retain only the data satisfying $0.05H < \Delta y < 0.2H$ {\color{black}in the fitting procedure}. This filtering eliminates the influence of strong nonlinear effects and surface imperfections{\color{black}. The upper bound of $0.2H$ is chosen to ensure optimal fitting across all parameter sets reported in Tables~\ref{tab:combined_critical_values}–\ref{tab:voltage_critical_values}, although in some cases (see Figs. \ref{fig:bif_r_1_5} and \ref{fig:fe_r_1_5_voltage}) the range of validity of Eq.~\eqref{eq:fitting} extends beyond this limit}. Eq.~\eqref{eq:fitting} allows to fit the data both in the case of a supercritical and subcritical transition. Indeed, the bifurcation is classified as supercritical if, in the filtered data, $\lambda < \lambda^\text{cr}$; otherwise, it is subcritical.
In Eq.~\eqref{eq:fitting}, $\hat A$ modulates the amplitude of the wrinkling close to the bifurcation point, while $\lambda^\text{num}_\text{cr}$ represents the numerical threshold of the wrinkling bifurcation.
The resulting fit demonstrates excellent agreement with the finite element results near the bifurcation point, see Fig.~\ref{fig:bif_r_1_5}(b).
Moreover, the numerically determined critical stretches $\lambda^\text{num}_\text{cr}$ closely match the thresholds $\lambda_\text{cr}$ predicted by the linearized stability analysis, with a precision of the order of $10^{-4}$, see Table~\ref{tab:combined_critical_values}. The only exception occurs for the case where $r=1/30$ and $\bar{E}_0=1$, where the discrepancy between the theoretical and numerical thresholds is of the order of $10^{-2}$. In this case, the bifurcation is subcritical and abrupt, becoming nonlinear very close to the bifurcation threshold and influencing the fitting procedure.

We note a turning point in all bifurcation diagrams in the fully nonlinear regime. Corresponding to these turning points, the finite element simulations reveal a progressive localization of the deformation, indicating that the strain becomes concentrated in narrow regions near the furrows of the wrinkling pattern rather than remaining uniformly distributed across the surface. This process eventually leads to self-contact (see the deformed configurations corresponding to points $A,\,B,\,C$ indicated in the bifurcation diagram in Fig.~\ref{fig:bif_r_1_5}), a transition reminiscent of the creasing onset observed by \cite{hohlfeld2011unfolding}.

\begin{figure}[t!]
    \centering
    \begin{minipage}{0.5\textwidth}\small(a)
    \end{minipage}\begin{minipage}{0.5\textwidth}
        \small(b)
    \end{minipage}
    \includegraphics[width=\textwidth]{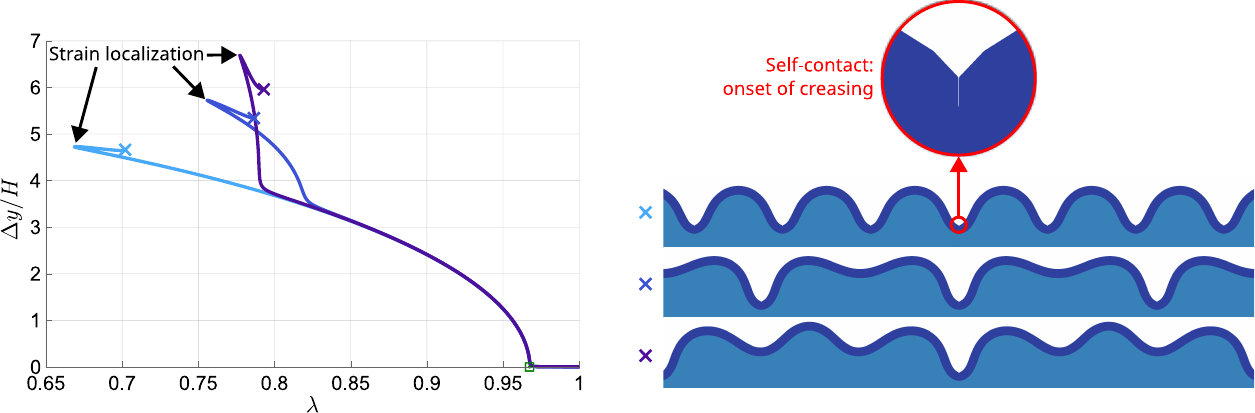}\caption{Results of the finite element simulations for $r=1/30$ and $\bar{E}_0 = 0.3$. {\color{black}(a) Non-dimensional} amplitude of the wrinkling of the free surface $\Delta y/H$ versus the stretch $\lambda$. The light blue line represents the wrinkling solution with a constant wavelength, while the blue and purple lines show the amplitude of the wrinkling pattern when period-doubling and period-tripling secondary bifurcations occur. The green square denotes the marginal stability threshold predicted by the linearized stability analysis, while the cross indicates when self-contact of the film occurs. {\color{black} The fitting of Eq.~\eqref{eq:fitting} is not shown here, as its range of validity is too limited compared to the large amplitude of the wrinkling pattern.} (b) Final morphologies of the finite element simulations at the onset of self-contact (represented in the inset), for the fixed-wavelength, period-doubling, and period-tripling solutions.}
    \label{fig:period_doub}
\end{figure}

Recall that film-substrate systems may also exhibit period-doubling and period-tripling secondary bifurcations, see the works by \cite{brau2011multiple}, \cite{cao2012Wrinkling}, \cite{fu2015asymptotic}, and \cite{budday2015period}.
For our soft dielectric film-substrate systems, we take a film 30 times stiffer than the substrate ($r=1/30$). To investigate secondary bifurcations, we conduct finite element simulations in computational domains that are twice and three times the fundamental length, respectively, while superposing imperfections corresponding to double and triple the critical wavelength. An intriguing feature of the system is the emergence of secondary bifurcations in the form of period-doubling and period-tripling (see Fig.~\ref{fig:period_doub}), whose occurrence is strongly influenced by the selected computational domain size and the characteristic length of the imposed imperfections. Again, we note turning points in the nonlinear regime. In these cases as well, the turning points correspond to a localization of the deformation close to the wrinkling furrows, which later evolve into self-contacting creases.

We also analyze the behavior of the system when $\lambda$ is held fixed and the applied non-dimensional voltage $\bar{E}_L$ is used to trigger the wrinkling instability.
It is observed from Fig.~\ref{fig:fe_r_1_5_voltage} that the system undergoes a supercritical transition at the onset of instability.
In contrast to the stretch-induced case, the finite element simulations in this scenario terminate before the onset of strain localization. We conjecture that this may be due to dielectric breakdown through catastrophic thinning~\citep{zurlo2017catastrophic}, a phenomenon associated with the loss of convexity in the energy functional, potentially leading to the non-existence of energy minimizers.

\begin{figure}[t!]
    \centering
    \includegraphics[width=0.6\textwidth]{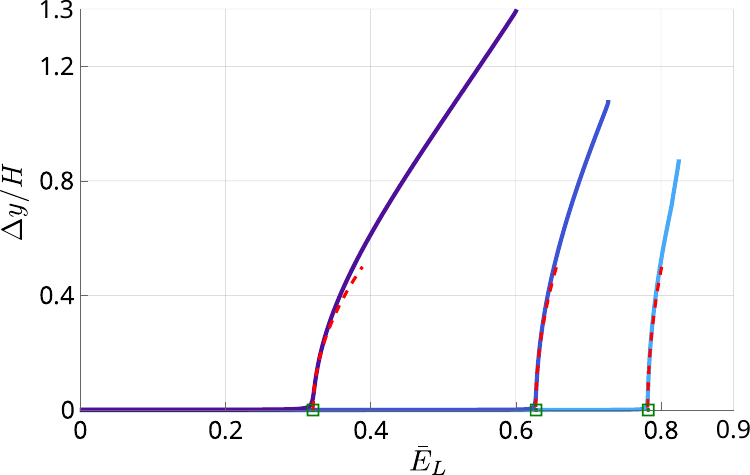}
    \caption{Bifurcation diagrams showing the non-dimensional wrinkling amplitude $\Delta y/H$ versus the applied non-dimensional voltage $\bar{E}_L$ for $r=1/5$ and $\lambda = 0.85,\,0.9,$ and $0.95$ (purple, blue, and light blue lines, respectively). The green squares denote the marginal stability thresholds obtained from the linearized stability analysis. The dashed lines show the best fit of the finite element data close to the bifurcation point with the function $\hat A \sqrt{\bar{E}_L-\bar{E}_\text{num}^\text{cr}}$ with $\hat A,\,\bar{E}_\text{num}^\text{cr}>0$, see the main text for details. Here, the fitted parameter $\hat A = 1.9198,\,2.9683,\, 3.6174$ and $\bar{E}_\text{num}^\text{cr}=0.3214,\,0.6278,\,0.7819$ for $\lambda = 0.85,\,0.9,\,0.95$, respectively.}
    \label{fig:fe_r_1_5_voltage}
\end{figure}

\begin{figure}[t!]
    \centering
    \begin{minipage}[t]{0.5\textwidth}
        \small(a) \\

        \vspace{0.5em}
        \includegraphics[width=\textwidth]{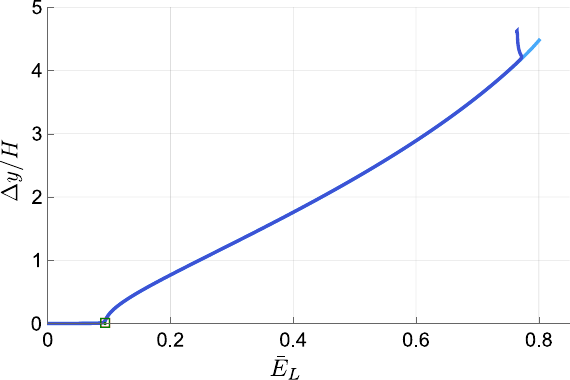}
    \end{minipage}\begin{minipage}[t]{0.5\textwidth}
        \small (b)\\
        \includegraphics[width=\textwidth]{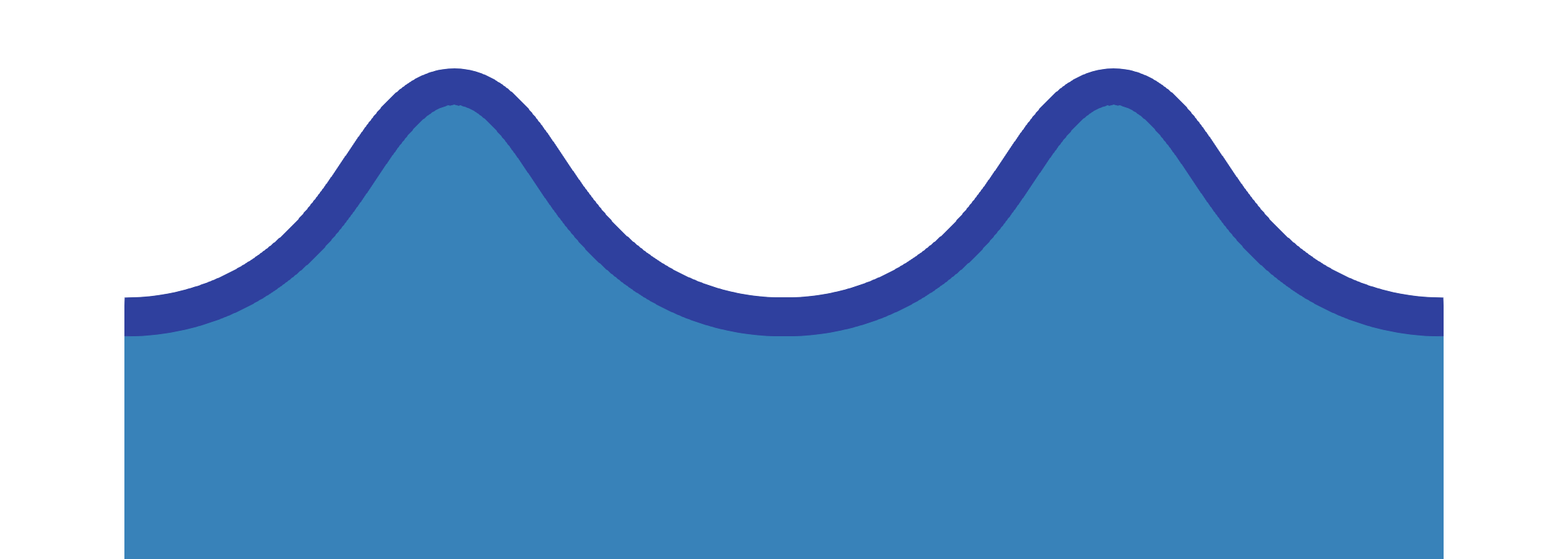}\\
        \includegraphics[width=\textwidth]{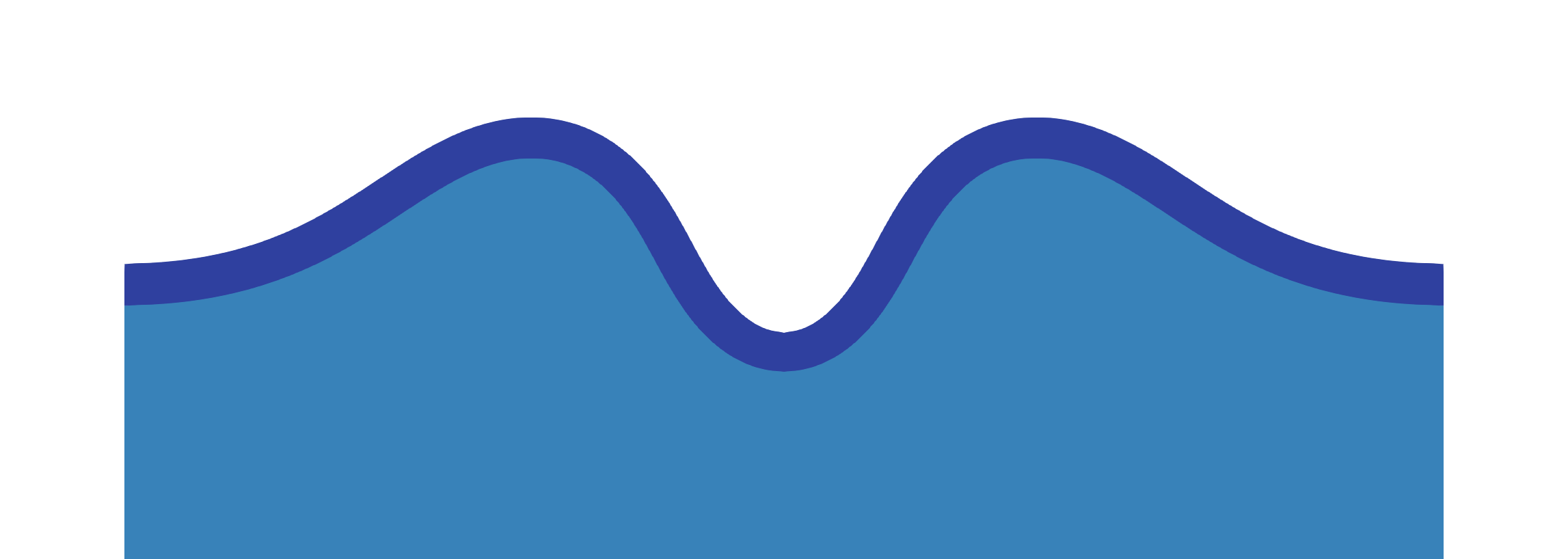}\\
    \end{minipage}
    \caption{(a) Bifurcation diagrams showing the non-dimensional wrinkling amplitude $\Delta y/H$ versus the applied non-dimensional voltage $\bar{E}_L$ for $r = 1/30$ and $\lambda = 0.95$. The green square denotes the marginal stability threshold obtained from the linearized stability analysis. The light blue line represents the wrinkling solution with a constant wavelength, while the blue line corresponds to the amplitude of the pattern after a period-doubling bifurcation occurs. {\color{black}The fitting of Eq.~\eqref{eq:fitting} is not explicitly shown here, as its range of validity is too limited compared to the large amplitude of the wrinkling pattern.} (b) Final morphology from the finite element simulations for both the fixed-wavelength and period-doubling solutions.}
    \label{fig:period_doub_voltage}
\end{figure}

As in the previous case, the bifurcation curves closely resemble those of supercritical pitchfork transitions. To verify this observation, we fit the finite element numerical data using the function $\hat A\sqrt{|\bar{E}_L - \bar{E}_\text{num}^\text{cr}|}$, following the same fitting procedure described earlier. The bifurcation is considered supercritical if $\bar{E}_L > \bar{E}_\text{num}^\text{cr}$ close to the bifurcation point, and otherwise subcritical.  The agreement is excellent near the bifurcation point, with the numerically predicted thresholds $\bar{E}_\text{num}^\text{cr}$ deviating from the theoretical values by less than 1\% in almost all cases, see Fig.~\ref{fig:fe_r_1_5_voltage} and Table~\ref{tab:voltage_critical_values} for a quantitative comparison with the theoretical thresholds. For the case $r=1/5$ and $\lambda=1.2$, the finite element simulations are highly sensitive to the imposed imperfection. To trigger the instability, we have to increase the imperfection amplitude to $1.5 \times 10^{-4}$ (compared to the baseline value of $5 \times 10^{-5}$, with all lengths scaled by the coating thickness). This adjustment causes the finite element numerical prediction of the critical voltage to appear at a lower value. Although decreasing the imperfection amplitude improves the accuracy of the critical threshold, in this case the simulations fail to converge once the bifurcation point is reached.

Furthermore, we observe period-doubling bifurcations when the soft dielectric film is sufficiently stiff relative to the substrate ($r = 1/30$), as illustrated in Fig.~\ref{fig:period_doub_voltage}(a). {\color{black}In the bifurcation diagram, we find excellent agreement with the linearized stability analysis. Similarly to the simulations shown in Fig.~\ref{fig:fe_r_1_5_voltage}, we do not observe any strain localization or self-contact. Instead, the resulting post-bifurcation morphology features ridges separated by elongated furrows, as displayed in Fig.~\ref{fig:period_doub_voltage}(b). Interestingly, a ridge morphology emerges here as a result of the instability for relatively large values of $r$ compared to purely elastic passive systems, where $ r < 10^{-3}$, see \cite{wang2015three}.}

\section{Conclusions}
\label{section7}

We presented a comprehensive theoretical analysis of the wrinkling instability of a soft dielectric film bonded to a hyperelastic substrate under the combined action of {\color{black}applied} voltage and plane-strain mechanical loading.

By relying on the Stroh formulation and the surface impedance matrix method, we {\color{black}obtained} exact bifurcation equations and accurate sixth-order approximate bifurcation equations.
We also derived explicit bifurcation equations of critical stretch $\lambda_\text{cr}$, voltage $\bar E_{L\text{cr}}$ and wavenumber $(kh)_\text{cr}$.
The asymptotic solution agrees well with the exact solution when $r$ is small, meeting the assumptions (of order ${\left( {kh} \right)^3}$ for small $kh$). Furthermore, we found that the thresholds of the shear modulus ratio {\color{black}$r_{\rm c}^0$ and pre-stretch $\lambda_{\rm c}^0$} for electro-elastic wrinkling {\color{black}correspond} to the purely mechanical instability case.

Finally, our finite element simulations further enriched these findings by exploring post-buckling behavior and complex pattern evolution beyond the initial wrinkle formation.
The simulations confirmed that the analytical critical points accurately mark the onset of instability, and revealed what happens beyond this point.
We observed secondary bifurcations such as period-doubling and tripling of the wrinkle pattern when the film is relatively stiff compared to the substrate. These secondary patterns imply that a single system can support multiple modes of surface morphology, which could be harnessed to achieve different functional states (for example, switching between two distinct wrinkle wavelengths under different electrical inputs).

The simulations also uncovered a limit to the tunability, because at high voltage levels (beyond a turning point in the bifurcation diagram), localized strain concentrations can lead to the formation of a sharp crease with self-contact. This incipient creasing is a critical consideration for applications, as it represents a material failure or extreme deformation state that designers might wish to avoid.

Thus, our results not only map out the desired regime for reversible wrinkling but also delineate the boundaries where the surface topology might become unstable in a destructive way.

\section*{Acknowledgements}

This work was supported by the National Natural Science Foundation of China (Nos.~12402107, 12192211, and U24A2005) and by the specialized research projects of Huanjiang Laboratory, Zhuji, Zhejiang Province, China. Additionally, this work was partially supported by the 111 Project, China (No.~B21034). The work of MD was supported by a Qiushi Guest Professor appointment (Seagull Program) at Zhejiang University. The work of DR was partially supported by INdAM through the project \emph{MATH-FRAC: MATHematical modelling of FRACture in nonlinear elastic materials}.

\appendix
\section{Exact and approximate bifurcation equations}
\label{AppeA}

The basic equations governing the finite electro-elastic deformations of an
incompressible soft electro-elastic body are well-established, and there is no need to repeat them here.
The same remark applies to the Stroh formulation of the equations of incremental deformations with sinusoidal variations along $x_1$ and exponential variations along $x_2$.
We refer the interested reader to the works of \cite{dorfmann2014nonlinear, su2018wrinkles, dorfmann2019electroelastic, broderick2020electro, su2020effect, yang2023tutorial}, for example.

As summari{\color{black}z}ed in Section \ref{Exact bifurcation}, the {\color{black}generalized, non-dimensional} displacement-traction vector $\bm{\eta}$ satisfies $\bm{\eta}' = \rm{i} \mathbf N \bm{\eta}$,  where {\color{black}$\rm{i}=\sqrt{-1}$ is the imaginary unit,} the prime denotes differentiation with respect to $kx_2$, and $\mathbf N$ is the (constant) Stroh matrix, which is {\color{black}partitioned} as $\mathbf{N} = \begin{bmatrix} \mathbf{N}_1 & \mathbf{N}_2 \\ \mathbf{N}_3 & \mathbf{N}_1^\text{T} \end{bmatrix}$.
For a general triaxial pre-stretch $(\lambda_1, \lambda_2, \lambda_3)$, we find that for the dielectric film {\color{black}characterized by the neo-Hookean ideal dielectric model \eqref{eq:energies}},
\begin{align}
     & {\mathbf N}_1
    = \begin{bmatrix}
          0  & -1 & 0 \\
          -1 & 0  & 0 \\
          0  & 0  & 0
      \end{bmatrix},
    \qquad
    {\mathbf N}_2
    = \begin{bmatrix}
          \lambda_1^2 \lambda_3^2         & 0 & \lambda_1^3\lambda_3^{3}\bar{E}_L     \\
          0                               & 0 & 0                                     \\
          \lambda_1^3\lambda_3^3\bar{E}_L & 0 & 1 + \lambda_1^4\lambda_3^4\bar{E}_L^2
      \end{bmatrix},
    \notag           \\[6pt]
     & {\mathbf N}_3
    = \begin{bmatrix}
          -(\lambda_1^2 + 3\lambda_1^{-2}\lambda_3^{-2}
          + 3\lambda_1^{2}\lambda_3^2\bar{E}_L^2)
           & 0
           & 2\lambda_1\lambda_3\bar{E}_L                                                       \\
          0
           & \lambda_1^{-2}\lambda_3^{-2} - \lambda_1^{2} + \lambda_1^{2}\lambda_3^2\bar{E}_L^2
           & 0                                                                                  \\
          2\lambda_1\lambda_3\bar{E}_L
           & 0
           & -1
      \end{bmatrix},
\end{align}
in our non-dimensional form, {\color{black}with $\bar E_L = \sqrt{\varepsilon/\mu_f}(V/H)$ representing the non-dimensional voltage} (for more general formulas, see \cite{su2018wrinkles}, where $\mathbf N$ is derived for a generic total free energy {\color{black}density function}).
The eigenvalues of $\mathbf N$ with positive imaginary parts are $q_1 = \mathrm i$, $q_2 = \mathrm i \lambda_1^2\lambda_3$, $q_3 = \mathrm i$, and the corresponding eigenvectors are the columns of the $6 \times 3$ matrix below,
\begin{equation}
    \begin{bmatrix} \boldsymbol{\eta}^{(1)} | \boldsymbol{\eta}^{(2)} | \boldsymbol{\eta}^{(3)} \end{bmatrix}
    =\begin{bmatrix}
        -\mathrm i \lambda_1^2\lambda_3^2          & -\mathrm i\lambda_1^4\lambda_3^3                                  & 0                                    \\
        \lambda_1^2\lambda_3^2                     & \lambda_1^2\lambda_3^2                                            & 0                                    \\
        - \mathrm i \lambda_1^3\lambda_3^3\bar E_L & -\mathrm i\lambda_1^5\lambda_3^4\bar E_L                          & \mathrm i                            \\
        2 + \lambda_1^4\lambda_3^4\bar E_L^2       & 1 + \lambda_1^4 \lambda_3^{2} + \lambda_1^4 \lambda_3^4\bar E_L^2 & \lambda_1\lambda_3\bar E_L           \\
        \mathrm i (1+ \lambda_1^4\lambda_3^2)      & 2\mathrm i\lambda_1^2 \lambda_3                                   & \mathrm i \lambda_1\lambda_3\bar E_L \\
        -\lambda_1^3\lambda_3^3\bar E_L            & - \lambda_1^3\lambda_3^3\bar E_L                                  & -1
    \end{bmatrix}.
    \label{e-vectors}
\end{equation}
The eigenvalues with negative imaginary parts are $q_4$, $q_5$,  $q_6$ with associated eigenvectors $\boldsymbol{\eta}^{(4)}$, $\boldsymbol{\eta}^{(5)}$, $\boldsymbol{\eta}^{(6)}$, which are the complex conjugates of $q_1$, $q_2$, $q_3$ and $\boldsymbol{\eta}^{(1)}$, $\boldsymbol{\eta}^{(2)}$, $\boldsymbol{\eta}^{(3)}$, respectively.
The $6 \times 6$ complete matrix of eigenvectors is defined as $\boldsymbol{\mathcal N}=\begin{bmatrix} \boldsymbol{\eta}^{(1)} | \boldsymbol{\eta}^{(2)} | \boldsymbol{\eta}^{(3)} | \boldsymbol{\eta}^{(4)} | \boldsymbol{\eta}^{(5)} | \boldsymbol{\eta}^{(6)} \end{bmatrix}$.

If the soft dielectric material were to occupy an entire half-space, then its impedance matrix would be $\mathbf Z = -\mathrm i \mathbf{BA}^{-1}$, where $\mathbf A$ and $\mathbf B$ are the $3 \times 3$ top and bottom submatrices of Eq.~\eqref{e-vectors}, respectively, or
\begin{equation}
    \mathbf{Z} =
    \begin{bmatrix}
        \lambda_{1}^{-2} \lambda_{3}^{-2 } + \lambda_{3}^{-1} +  \lambda_{1}^2 \lambda_{3}^2 \bar{E}_{L}^2
                                              & - \mathrm i (\lambda_{1}^{-2} \lambda_{3}^{-2 } - \lambda_{3}^{-1} +  \lambda_{1}^2 \lambda_{3}^2 \bar{E}_{L}^2 )
                                              & - \lambda_{1} \lambda_{3} \bar{E}_{L}                                                                                                                               \\
        \mathrm i (\lambda_{1}^{-2} \lambda_{3}^{-2 } - \lambda_{3}^{-1} +  \lambda_{1}^2 \lambda_{3}^2 \bar{E}_{L}^2 )
                                              & \lambda_{1}^{2} + \lambda_{3}^{-1}                                                                                & - \mathrm i \lambda_{1} \lambda_{3} \bar{E}_{L} \\
        - \lambda_{1} \lambda_{3} \bar{E}_{L} & \mathrm i \lambda_{1} \lambda_{3} \bar{E}_{L}                                                                     & 1
    \end{bmatrix}.
    \label{Z-general}
\end{equation}
The bifurcation condition for {\color{black}the Biot-type surface instability} would then be: $\det \mathbf Z = 0$ \citep{destrade2008surface, destrade2015incremental}, or
\begin{equation} \label{A.4}
    \lambda_{1}^{6}\lambda_{3}^{3}
    +\lambda_{1}^{4}\lambda_{3}^{2}
    +3\,\lambda_{1}^{2}\lambda_{3}
    -1
    = \lambda_{1}^{4}\lambda_{3}^{4}(1 + \lambda_{1}^{2}\lambda_{3}) \bar E_{L}^{2}.
\end{equation}
In plane strain ($\lambda_1=\lambda$, $\lambda_3=1$), the bifurcation equation \eqref{A.4} reduces to Eq.~\eqref{surf-inst-plane}, while in equi-biaxial strain ($\lambda_1=\lambda_3=\lambda$), it recovers the formula established by \cite{su2018wrinkles}.

Here, however, the soft dielectric film has a finite thickness and is in contact with the elastic substrate.
The $3 \times 1$ generalized, non-dimensional traction {\color{black}$\mathbf S=\left[S_{21}, S_{22}, \Phi \right]^\text{T}$ and displacement $\mathbf U=\left[ U_1, U_2, \Delta \right]^\text{T}$} vectors on each side of the interface at $x_2=0$ are related through
\begin{equation} \label{A.5}
    \mathbf S_f(0)=  \mathrm i \mathbf Z_f\mathbf U_f(0), \qquad \mathbf S_s(0) =  \mathrm i \mathbf Z_s \mathbf U_s(0),
\end{equation}
so that the boundary conditions of perfect bond at the interface, $\mathbf U_f(0) =  \mathbf U_s(0)$ and $\mu_f \mathbf S_f(0)=\mu_s \mathbf S_s(0)$, yields the bifurcation condition as Eq.~\eqref{3.3.9}: $\det(\mathbf Z_f - r \mathbf Z_s) = 0$ (see e.g., \cite{shuvalov2002some}).
Here, the film impedance matrix $\mathbf Z_f$ at $x_2=0$,  assuming the $x_2=-h$ surface is traction-free and the applied voltage remains constant, is defined as $\mathbf Z_f = -\mathrm i \mathbf M_3 \mathbf M_1^{-1}$, where $\mathbf M_1$ and $\mathbf M_3$ are, respectively, the $3 \times 3$ upper-diagonal and lower-off-diagonal submatrices of the $6 \times 6$ {\color{black}exponential} matrix $\mathbf M = \exp(\textrm i kh \mathbf N)$, which can be computed as $\mathbf M =  \bm{\mathcal N} \mathbf \Lambda \bm{\mathcal N}^{-1}$, with $\mathbf \Lambda$ the diagonal matrix with elements $e^{\mathrm i q_jkh}$ ($j=1,\ldots,6$).
The substrate impedance matrix $\mathbf Z_s$ reads as follows,
\begin{equation}
    \mathbf{Z}_s =
    \begin{bmatrix}
        \lambda_{1}^{-2} \lambda_{3}^{-2 } + \lambda_{3}^{-1}
          & - \mathrm i (\lambda_{1}^{-2} \lambda_{3}^{-2 } - \lambda_{3}^{-1})
          & 0
        \\
        \mathrm i (\lambda_{1}^{-2} \lambda_{3}^{-2 } - \lambda_{3}^{-1})
          & \lambda_{1}^{2} + \lambda_{3}^{-1}                                  & 0 \\
        0 & 0                                                                   & 0
    \end{bmatrix},
\end{equation}
which is consistent with Eq.~\eqref{Z-general} written at $\bar E_L=0$, provided the last diagonal entry there is replaced with a zero to account for the two-dimensional nature of the traction and displacement vectors in the hyperelastic substrate (with no electric field).

We can solve the exact bifurcation equation \eqref{3.3.9} numerically, but it can prove computationally costly, which is why we may wish to use small-parameter expansions and conduct asymptotic analysis.

    {\color{black}The solution of the first-order differential equation, $\bm{\eta}' = \rm{i} \bf N \bm{\eta}$, with a constant Stroh matrix $\mathbf N$, is $\bm \eta(k x_2) = \exp (\text i k x_2 \mathbf N) \bm \eta(0)$. Therefore, the relationship between the generalized displacement-traction vectors of the upper and lower surfaces of the soft dielectric film reads $\bm \eta(-kh) = \exp (- \text i kh \mathbf N) \bm \eta(0) \equiv \mathbf {\hat M} \bm \eta(0)$. From the continuity conditions of $\bm \eta$ at the interface $x_2=0$ ($\mathbf U_f(0) =  \mathbf U_s(0)$ and $\mathbf S_f(0)=r \mathbf S_s(0)$), and the conditions of zero traction and a constant applied voltage on the top surface $x_2=-h$ ($\mathbf S_f(-kh)=\bm 0$), it follows that
        \begin{equation} \label{A.7.new}
            \left[ {
                        \begin{array}{*{10}{c}}
                            {{\bf{U}}_f\left( { - kh} \right)} \\
                            {\bf{0}}
                        \end{array}} \right] = {\bf{\hat M}}\left[ {
                        \begin{array}{*{20}{c}}
                            {{{\bf{U}}_{s}}(0)} \\
                            {r{{\bf{S}}_{s}}(0)}
                        \end{array}} \right].
        \end{equation}}

For thin dielectric films,  where $kh \ll 1$, {\color{black}we substitute \eqref{A.5}$_2$ into \eqref{A.7.new} and write the power series $\mathbf {\hat M} \equiv \exp(-\textrm i kh \mathbf N) =  \sum \frac{1}{n!}(- \mathrm i \mathbf N)^n(kh)^n$} to arrive at the sixth-order approximation of the exact bifurcation equation,
\begin{equation} \label{4.2.1}
    \det \left[
        \begin{array}{l}
            \mathrm i r \mathbf Z_s + (r \mathbf N_1 \mathbf Z_s - \mathrm i  \mathbf N_3)(kh)
            - \frac{1}{2}\mathrm i (r \mathbf K^{(2)}_4 \mathbf Z_s - \mathrm i \mathbf K^{(2)}_3)(kh)^2 \\
            - \frac{1}{6}(r \mathbf K^{(3)}_4 \mathbf Z_s - \mathrm i \mathbf K^{(3)}_3)(kh)^3
            + \frac{1}{24}\mathrm i(r \mathbf K^{(4)}_4 \mathbf Z_s - \mathrm i \mathbf K^{(4)}_3)(kh)^4 \\
            + \frac{1}{120}(r\mathbf K^{(5)}_4 \mathbf Z_s - \mathrm i \mathbf K^{(5)}_3)(kh)^5
            - \frac{1}{720}\mathrm i (r \mathbf K^{(6)}_4 \mathbf Z_s - \mathrm i \mathbf K^{(6)}_3)(kh)^6
        \end{array} \right] = 0,
\end{equation}
where $\mathbf K^{(n)}_3$ and $\mathbf K^{(n)}_4$ are, respectively, the $3 \times 3$ lower-off-diagonal and lower-diagonal submatrices of the $6 \times 6$ matrix $\mathbf K^{(n)} \equiv \mathbf N^n$.

Solving this approximate bifurcation equation \eqref{4.2.1} numerically is much more efficient and less computationally expensive than solving the exact bifurcation condition \eqref{3.3.9}, and it is highly accurate for small $kh$ and small $r$.

{\color{black}\section{Asymptotic analysis of approximate bifurcation equation \eqref{4.2.1}}
\label{AppeB}

Following \cite{Cai_2000Exact}, we can use the approximate bifurcation equation \eqref{4.2.1} in this appendix to derive power-series asymptotic expansions in $kh$ for the stretch $\lambda$ and voltage $\bar E_{L}$ when the soft dielectric film is much stiffer than the substrate, and further, explicit asymptotic expansions of the critical values $\lambda_\text{cr}$ and $\bar E_L^\text{cr}$ in powers of $r^{1/3}$.
Here we focus on the plane-strain loading case.

Assuming $kh \ll 1$ and $r$ of order $(kh)^3$, an expansion of the sixth-order approximate bifurcation condition \eqref{4.2.1}, followed by elimination of the common factor, leads to
\begin{multline} \label{B1}
    {\omega _0}+ {\omega _1}\left( {kh} \right) + \tfrac{1}{2}{\omega _2}{\left( {kh} \right)^2} + \tfrac{1}{6}{\omega _3}{\left( {kh} \right)^3} + \tfrac{1}{{24}}{\omega _4}{\left( {kh} \right)^4} \\[4pt]
    + \tfrac{1}{{120}}{\omega _5}{\left( {kh} \right)^5} + \tfrac{1}{{720}}{\omega _6}{\left( {kh} \right)^6}+{\cal O}\left( {\left({kh}\right)^7} \right) = 0,
\end{multline}
where
\begin{equation}\label{B2}
    \begin{array}{l}
        {\omega _0} = \left( { - 1 + 3{\lambda ^2} + {\lambda ^4} + {\lambda ^6}} \right){r^2},                                                                                                \\[4pt]
        {\omega _1} =  - \left( {1 + {\lambda ^2}} \right)\left[ {1 - 3{\lambda ^2} + \left( {\bar E_L^2 - 1} \right) {\lambda ^4} \left( {1 + {\lambda ^2}} \right)} \right] r,               \\[4pt]
        {\omega _2} = 2\left( {\bar E_L^2 - 1} \right) \left[ {\left( {\bar E_L^2 - 1} \right){\lambda ^4}} - 2 \right] {\lambda ^4} - 6 - 8\left( {{\lambda ^2} - 1} \right)r,                \\[4pt]
        {\omega _3} =  - \left( {1 + {\lambda ^2}} \right)\{ 2 - 4{\lambda ^2} + {\lambda ^4}[ - 7 - 2\bar E_L^4{\lambda ^4} - {\lambda ^2}\left( {7 + 3{\lambda ^2} + {\lambda ^4}} \right)   \\[4pt]
        \qquad \;  + \bar E_L^2\left( {6 + 4{\lambda ^2} + 5{\lambda ^4} + {\lambda ^6}} \right)] \} r,                                                                                        \\[4pt]
        {\omega _4} =  - 16 + 4{\lambda ^4} \left[{5 - 2 \bar E_L^2 + 2\left( {3 - 4\bar E_L^2 + \bar E_L^4} \right){\lambda ^4} + {{\left( {\bar E_L^2 - 1} \right)}^2}{\lambda ^8}} \right], \\[4pt]
        {\omega _5} =  f(\lambda,\bar E_L^2) r,
        \\[4pt]
        {\omega _6} =  - 40 - 8\left( {\bar{E}_L^2 - 9} \right){\lambda ^4} + 2\left( {91 - 92\bar{E}_L^2 + 16\bar{E}_L^4} \right){\lambda ^8}                                                 \\[4pt]
        \qquad \; + 4\left( {17 - 27\bar{E}_L^2 + 10\bar{E}_L^4} \right){\lambda ^{12}} + 6{\left( {\bar{E}_L^2 - 1} \right)^2}{\lambda ^{16}},
    \end{array}
\end{equation}
where we omit the explicit form of $f(\lambda, \bar E_L^2)$ for brevity.
Because
$\omega_5$ depends linearly on $r$, the sixth term in Eq.~\eqref{B1} is of order $(kh)^8$ and may consequently be discarded.

\subsection{Critical stretch under a prescribed electric voltage} \label{AppeB.1}

For a prescribed non-dimensional electric voltage $\bar E_L=\bar E_0$, we first derive a power-series asymptotic expansion of the stretch $\lambda$ in $kh$, from which the explicit asymptotic expansion of the critical stretch $\lambda_\text{cr}$ in powers of $r^{1/3}$ follows.

As $r = {\cal O}\left({(kh)^3} \right)$, the leading-order term arises from the third term in Eq.~\eqref{B1}, which is of order $(kh)^2$. Thus, Eq.~\eqref{B1} reduces to the leading-order bifurcation condition,
\begin{equation}\label{B3}
    3 - \left( {1 - \bar E_0^2} \right) \lambda^4 \left[ {2 + \left( {1 - \bar E_0^2} \right){\lambda ^4}} \right] = 0,
\end{equation}
which gives the leading-order expression for the (critical) stretch, $\lambda_0 = (1-\bar E_0^2)^{-1/4}$, as presented in Eq.~\eqref{initial}. %

Examination of Eq.~\eqref{B1} reveals that the coefficient $\phi_1$ in the first-order asymptotic expansion $\lambda = \lambda_0 + \phi_1 kh$ vanishes, and that the next-order expansion is $\lambda = \lambda_0 + \phi_2 (kh)^2$.
Substituting this into Eq.~\eqref{B1} and equating the coefficients of $(kh)^4$ yields
\begin{equation}\label{B4}
    1 + 3\left( {1 + \lambda _0^{ - 2} } \right) r/(kh)^3 + 12{ \lambda _0^{ - 5}}\phi_2 = 0,
\end{equation}
which gives ${\phi _2}$, and the second-order correction to the stretch as
\begin{equation}\label{B5}
    \lambda = \lambda _0   - \frac{1}{4}(\lambda_0^3+\lambda_0^5) \left(\frac{r}{kh}\right)  - \frac{1}{12}\lambda_0^5 (kh)^2.
\end{equation}

In a similar manner, substituting the third- and fourth-order asymptotic expansions, $\lambda = \lambda_0 + \phi_2 (kh)^2 + \phi_3 (kh)^3$ and $\lambda = \lambda_0 + \phi_2 (kh)^2 + \phi_3 (kh)^3 + \phi_4 (kh)^4$, into Eq.~\eqref{B1} and equating the coefficients of $(kh)^5$ and $(kh)^6$ yields $\phi_3$ and $\phi_4$, respectively. Their explicit forms are omitted here for brevity. The resulting fourth-order asymptotic expansion of the stretch $\lambda$ is presented in Eq.~\eqref{4.2.2}.

Subsequently, we determine, in turn, the critical wavenumber $(kh)_{\rm cr}$ in Eq.~\eqref{4.2.5} and the critical stretch $\lambda_{\rm cr}$ in Eq.~\eqref{4.2.4} by setting the derivative of Eq.~\eqref{4.2.2} with respect to $kh$ equal to zero.

\subsection{Critical electric voltage for a fixed pre-stretch}
\label{AppeB.2}

Here we derive a power-series asymptotic expansion in $kh$ for ${\bar E}_L^2$ at a fixed pre-stretch $\lambda$, from which the explicit asymptotic expansion of the squared critical voltage $({\bar E}_L^{\rm cr})^2$ in powers of $r^{1/3}$ can be obtained.

Analogous to the derivation of the asymptotic expansion of the stretch $\lambda$ presented in \ref{AppeB.1}, the leading-order term of the squared voltage is obtained from Eq.~\eqref{B3} as $\bar E_{L0}^2 = 1 - \lambda^{-4}$. In a similar manner, the second-order asymptotic expansion of the squared voltage can be derived as
\begin{equation}\label{B2-1}
    \bar E_L^2  =\bar E_{L0}^2
    + \frac{1}{3}(kh)^2
    + (1 + \lambda^{-2})\left(\frac{r}{kh} \right),
\end{equation}
and the resulting fourth-order asymptotic expansion of $\bar E_L^2$ is formulated in Eq.~\eqref{4.2.3}.

By subsequently setting the derivative of Eq.~\eqref{4.2.3} with respect to $kh$ to zero, the asymptotic expansions of the critical wavenumber $(kh)^{\rm cr}$ and the critical squared voltage $({\bar E}_L^{\rm cr})^2$ in powers of $r^{1/3}$ are obtained, as given in Eqs.~\eqref{19} and \eqref{4.2.7}, respectively.}
\bibliographystyle{apalike}
\bibliography{refs.bib}

\end{document}

%% file: packages.tex
\usepackage[authoryear]{natbib}
\usepackage{graphicx} %
\usepackage{bm}
\usepackage{mathtools}
\usepackage[hidelinks]{hyperref}
\usepackage{xcolor}
\usepackage{xargs}
\usepackage{amsmath, amssymb, amsthm, cleveref}
\usepackage{subfig}
\usepackage{mathrsfs}
\usepackage{fullpage}
\graphicspath{{./Images/}}
\usepackage{booktabs}

%% file: defs.tex
\DeclareMathOperator{\Grad}{Grad}
\DeclareMathOperator{\tr}{tr}
\let\det\relax
\DeclareMathOperator{\det}{det}
\newcommand{\vect}[1]{\boldsymbol{#1}}
\newcommand{\tens}[1]{\mathsf{#1}}